\documentclass[12pt]{iopart}
\usepackage{iopams}
\usepackage{setstack}
\usepackage{graphicx,epstopdf}

\usepackage[colorlinks,linktocpage,linkcolor=blue]{hyperref}

\begin{document}

\title[]{Novel anomalous diffusion phenomena of underdamped Langevin equation with random parameters}

\author{Yao Chen$^1$ and Xudong Wang$^2$}

\address{$^1$College of Sciences, Nanjing Agricultural University, Nanjing, 210094, P.R. China  \\
$^2$School of Mathematics and Statistics, Nanjing University of Science and Technology, Nanjing, 210094, P.R. China}
\ead{ychen@njau.edu.cn and xdwang14@njust.edu.cn}
\vspace{10pt}

\begin{abstract}
The diffusion behavior of particles moving in complex heterogeneous environment is a very topical issue. We characterize particle's trajectory via an underdamped Langevin system driven by a Gaussian white noise with a time dependent diffusivity of velocity, together with a random relaxation timescale $\tau$ to parameterize the effect of complex medium. We mainly concern how the random parameter $\tau$ influences the diffusion behavior and ergodic property of this Langevin system. Besides, the comparison between the fixed and random initial velocity $v_0$ is conducted to show the effect of different initial ensembles. The heavy-tailed distribution of $\tau$ with finite mean is found to suppress the decay rate of the velocity correlation function and promote the diffusion behavior, playing a competition role to the time dependent diffusivity. More interestingly, a random $v_0$ with a specific distribution depending on random $\tau$ also enhances the diffusion. Both the random parameters $\tau$ and $v_0$ influence the dynamics of the Langevin system in an non-obvious way, which cannot be ignored even they has finite moments.
\end{abstract}

\section{Introduction}

Anomalous diffusion phenomenon is very common in both living and abiotic systems \cite{TruskeyYuanKatz:2004,BarkaiGariniMetzler:2012,HoflingFranosch:2013,Manzo:2015,MogreBrownKoslover:2020}. It is characterized by the significant deviation from Brownian motion with the nonlinear time dependence of the ensemble-averaged mean-squared displacement (EAMSD) \cite{MetzlerKlafter:2000,KlagesRadonsSokolov:2008}:
\begin{equation}
	\langle x^2(t)\rangle\propto t^\mu, ~~\mu\neq1.
\end{equation}
Different range of the anomalous exponent $\mu$ determines a more detailed division of the anomalous diffusion phenomenon: $\mu<1$ representing the subdiffusion and $\mu>1$ the superdiffusion.
The subdiffusion phenomena of passive particles abound in the cytoplasm of living biological cells \cite{WeberSpakowitzTheriot:2010,WeigelSimonTamkunKrapf:2011,Rienzo-etal:2014} and in artificially crowded fluids \cite{BanksFradin:2005,SzymanskiWeiss:2009,JeonLeijnseOddershedeMetzler:2013}, as well as in quasi-two-dimensional systems such as lipid bilayer membranes \cite{StachuraKneller:2015,JeonMonneJavanainenMetzler:2012,Jeon-etal:2016,AkimotoYamamotoYasuokaHiranoYasui:2011}. The superdiffusion is typically associated with active matter and can also be observed in living cells \cite{CaspiGranekElbaum:2000,RobertNguyenGalletWilhelm:2010,Reverey-etal:2015}.
Modern microscopy techniques have made it possible to track the random motions of tracer particles in a variety of living and abiotic systems.
The universal existence of the anomalous diffusion phenomena in the natural world inspires the scientists to devise the possible physical models and explore the underlying mechanisms of them.

The most typical model of describing particle's motion is continuous-time random walk (CTRW) \cite{MetzlerKlafter:2000,MontrollWeiss:1965,KlafterSokolov:2011,ZaburdaevDenisovKlafter:2015}, which is characterized by waiting time and jump length. By assuming the specific distributions of waiting time and jump length, together with the possible correlation between them, the CTRW can yield many kinds of diffusion processes, which present anomalous diffusion phenomena with different anomalous exponents \cite{ChechkinHofmannSokolov:2009,ChenWangDeng:2019,TejedorMetzler:2010,MagdziarzMetzlerSzczotkaZebrowski:2012-1,MagdziarzMetzlerSzczotkaZebrowski:2012-2}.
However, more and more new classes of diffusive dynamics have been observed, especially in the field of soft matter, biological, and other complex systems. One typical example is the Brownian non-Gaussian phenomenon, where the EAMSD is normal but the probability density function is non-Gaussian \cite{WangAnthonyBaeGranick:2009,ToyotaHeadSchmidtMizuno:2011,SilvaStuhrmannBetzKoenderink:2014,Bhattacharya-etal:2013,SamantaChakrabarti:2016}. The effective idea of interpreting this novel phenomenon is to introduce a random diffusivity which is analogous to the concept of superstatistics \cite{Beck:2001,BeckCohen:2003,Beck:2006}. It says that each one of particles moving in a complex heterogeneous environment has its own diffusivity \cite{WangKuoBaeGranick:2012,HapcaCrawfordYoung:2009,ChubynskySlater:2014,ChechkinSenoMetzlerSokolov:2017}. In general, the idea of superstatistics has been applied to many dynamics: turbulent dispersion \cite{Beck:2001}, renewal critical events in intermittent systems \cite{ParadisiKaniadakisScarfone:2015,AkinParadisiGrigolini:2009}, and different effective statistical mechanics with fluctuating intensive quantities \cite{Beck:2001,BeckCohen:2003}.

This paper considers an underdamped Langevin system driven by a Gaussian white noise with a time dependent diffusivity of velocity. The effect of the complex medium on particles is parameterized via a random relaxation timescale $\tau$. Each particle has its own fixed relaxation timescale $\tau$ as it moves in the complex medium, but the $\tau$ might be discrepant between different particles.
For the classical Brownian motion with a constant relaxation timescale $\tau$, the EAMSD evolves from ballistic diffusion to normal diffusion, with the crossover point at time $\tau$. One can imagine that a random relaxation timescale $\tau$ might yield novel dynamics especially when $\tau$ has a heavy-tailed distribution. On the other hand, to maintain the good statistics of the Langevin system with constant $\tau$, we suppose that the first moment $\langle\tau\rangle$ is finite. Considering the two conditions, the relaxation timescale $\tau$ is assumed to obey the modified L\'{e}vy distribution as (\ref{PDF}) shows.

We mainly concern the ergodic property of the underdamped Langevin system, which is embodied by the time-averaged mean-squared displacement (TAMSD), defined as  \cite{MetzlerJeonCherstvyBarkai:2014,BurovJeonMetzlerBarkai:2011,AkimotoCherstvyMetzler:2018}
\begin{equation}\label{TADef}
		\overline{\delta^2(\Delta)}
		=\frac{1}{T-\Delta}\int_0^{T-\Delta}  (x(t+\Delta)-x(t))^2{\rm d}t.
\end{equation}
Here,  $\Delta$ is the lag time, $T$ is the total measurement time.  To obtain good statistical properties, the lag time is assumed to be much shorter than the measurement time, i.e., $\Delta\ll T$. Compared with EAMSD, the TAMSD places emphasis on particle-to-particle diffusion properties and contains more information of particle's motion in a single trajectory. Based on the advance of single-particle tracking techniques, scientists often evaluate the recorded time series in terms of TAMSD to study the diffusion behavior of particles in living cells \cite{WeberSpakowitzTheriot:2010,GoldingCox:2006,BronsteinIsraelKeptenMaiTalBarkaiGarini:2009}.

A stochastic process is called ergodic if its TAMSD and EAMSD are equivalent, i.e., $\overline{\delta^2(\Delta)}=\langle x^2(\Delta)\rangle$ as the measurement time $T\rightarrow\infty$, such as Brownian motion and (tempered) fractional Brownian motion \cite{Goychuk:2012,DengBarkai:2009,ChenWangDeng:2017}.
But for anomalous diffusion processes, especially for the molecules diffusing in living cells, the time average often becomes a random variable and irreproducible for the individual trajectories, implying the ergodicity breaking, such as the continuous-time random walk \cite{WeigelSimonTamkunKrapf:2011,BelBarkai:2005_1,HeBurovMetzlerBarkai:2008}, L\'{e}vy walk \cite{FroembergBarkai:2013,GodecMetzler:2013,AlbersRadons:2018}, heterogeneous diffusion process \cite{CherstvyChechkinMetzler:2013,CherstvyMetzler:2013,CherstvyMetzler:2014,CherstvyMetzler:2015}, and random diffusivity process \cite{UneyamaMiyaguchiAkimoto:2015,Wang-etal:2020,WangChen:2021}, etc.

When investigating the ergodic property of a diffusion process, the effect of the initial condition usually cannot be ignored. The time average is in some sense an equilibrium measure, while the ensemble average is not. The different diffusion behavior resulting from the discrepant initial ensemble has been observed in the context of single file diffusion \cite{LeibovichBarkai:2013}, diffusing diffusivity model \cite{HidalgoBarkaiBurov:2021}, L\'{e}vy walk \cite{FroembergSchmiedebergBarkaiZaburdaev:2015,WangChenDeng:2019,WangChenDeng:2019-2}, biased random walk \cite{AkimotoCherstvyMetzler:2018,HouCherstvyMetzlerAkimoto:2018}. For the underdamped Langevin equation with random relaxation timescale $\tau$, we also study the effects of different initial velocity $v_0$, which is also the main part of this paper.

The key of evaluating the EAMSD and TAMSD is to obtain the velocity correlation function first, which has been deeply influenced by the random relaxation timescale $\tau$. Hence, the diffusion behavior and the ergodic property are closely related to the parameter $\tau$. Fixing the initial velocity $v_0$, we find that the random $\tau$ enhances the diffusion behavior while the time-dependent diffusivity slows down the diffusion behavior. They play the competition roles and present different diffusion phenomena.
On the other hand, when the initial velocity becomes random, especially depends on the random parameter $\tau$, the random initial velocity also enhances the diffusion behavior as the random relaxation timescale $\tau$.

The structure of this paper is as follows. In section \ref{Sec1},  we establish the underdamped Langevin system driven by the Gaussian white noise with a time dependent diffusivity. Then we evaluate the EAMSD and TAMSD, and make comparison between the cases with fixed and random relaxation timescale $\tau$ in section \ref{Sec2}, and between the cases with  fixed and random initial velocity $v_0$ in section \ref{Sec3}. Some discussions and summaries are provided in section \ref{Sec4}.
For readability, some mathematical details are present in Appendix.

\section{Model}\label{Sec1}

The underdamped Langevin equation driven by a Gaussian white noise with a time-dependent diffusivity is written as
\begin{equation}\label{langevin}
 \dot{x}(t)=v(t),~~~~~~
    \dot{v}(t)=-\tau^{-1}v(t)+\sqrt{2\nu t^{\beta-1}}\xi(t),
\end{equation}
where $\tau$ is the relaxation timescale, $\nu t^{\beta-1}$ is the time-dependent diffusivity with $0<\beta<1$, $\xi(t)$ is Gaussian white noise with zero mean value and correlation $\langle \xi(t_1)\xi(t_2) \rangle=\delta(t_1-t_2)$. This model can be regarded as the underdamped form of scaled Brownian motion, which describes the motion of a particle in a viscous medium with time-dependent temperature. Another kind of underdamped scaled Brownian motion described by  Langevin equation with internal noise is studied in \cite{Bodrova_etc:2016}, where the time dependent diffusivity is related to the damping coefficient according to the (time local) fluctuation-dissipation theorem \cite{Kubo:1966}.

The novelty of the Langevin equation (\ref{langevin}) is that the relaxation timescale $\tau$ is not a constant any more, but a random variable, due to the complexity of the media. Each particle has its own relaxation timescale $\tau$, which might discrepant for different particles. That is to say, the parameter $\tau$ is random for ensemble of particles, but fixed for each particle or trajectory. Our aim is to find how the random parameter $\tau$ influence the diffusion behavior and ergodicity of the Langevin system (\ref{langevin}) by evaluating the EAMSD and TAMSD of particles.
Besides, the initial velocity $v_0$ of particles can also be random, which might lead to a different phenomenon of the ergodicity from the cases with a fixed $v_0$.

Throughout this paper, we consider the relaxation timescale $\tau$ obeying the distribution \cite{Vitali_etc:2018}
\begin{equation}\label{PDF}		g(\tau)=\frac{\alpha}{\Gamma(1/\alpha)}\frac{1}{\tau}L_\alpha\left(\frac{\alpha}{\Gamma(1/\alpha)}\frac{\tau}{\langle\tau\rangle}\right),
\end{equation}
where $L_\alpha(\cdot)$ is the one-sided fully skewed $\alpha$-stable L\'{e}vy distribution with $0<\alpha<1$ \cite{KlafterSokolov:2011,Applebaum:2009} and $\langle\tau\rangle$ is the mean value of the random parameter $\tau$. The advantage of choosing such a kind of distribution is utilizing its heavy-tailed property $g(\tau)\propto \tau^{-2-\alpha}$ for large $\tau$. This property can yield a very large value of variable $\tau$, which may change the diffusion behavior of the Langevin equation (\ref{langevin}). The prefactor $1/\tau$ in front of the L\'{e}vy distribution is aimed to guarantee a finite mean $\langle\tau\rangle$ of the relaxation timescale. Otherwise, the variance of velocity would be infinite \cite{Vitali_etc:2018}. When simulating the random variable $\tau$, the parameter $\langle\tau\rangle$ controls the width of the L\'{e}vy distribution and can be chosen arbitrary.
For completeness on the description of the distribution in  (\ref{PDF}), we present its normalization and the mean value $\langle\tau\rangle$, together with the algorithm of generating $\tau$ in \ref{App-1}.

The key of obtaining the EAMSD and TAMSD is the velocity correlation function $\langle v(t_1)v(t_2)\rangle$. To obtain it, we firstly solve (\ref{langevin}) in Laplace domain, and achieve the analytic expression of velocity process
\begin{equation}
   v(t)=v_0{\rm e}^{-\frac{t}{\tau}}+\sqrt{2\nu}\int_0^t {\rm e}^{-\frac{t-t'}{\tau}}t'^{\frac{\beta-1}{2}}\xi(t'){\rm d}t',
\end{equation}
which is valid for both fixed and random $v_0$ and $\tau$. When $v_0$ and $\tau$ are random variables, instead of $\langle v(t_1)v(t_2)\rangle$, we express the velocity correlation function  in the condition of parameters $\tau$ and $v_0$, i.e., the conditional velocity correlation function
\begin{eqnarray}\label{covar}
    \langle v(t_1)v(t_2)|v_0,\tau\rangle
    &={\rm e}^{-\frac{t_1+t_2}{\tau}}\left(v^2_0+2\nu \int_0^{t_1}{\rm e}^{\frac{2t'}{\tau}}t'^{\beta-1} {\rm d}t'\right) \nonumber\\
    &={\rm e}^{-\frac{t_1+t_2}{\tau}}\left(v^2_0+\frac{2\nu t^{\beta}_1}{\beta} {_1F_1\left(\beta, 1+\beta, \frac{2t_1}{\tau}\right)}\right)
\end{eqnarray}
for $t_1\leq t_2$, where the ensemble average is only performed over the Gaussian white noise $\xi(t)$, and $_1F_1(a, b, z)$ is the confluent hypergeometric function.

\section{Fixed and random  relaxation timescale $\tau$}\label{Sec2}

\begin{figure}[!htb]
\flushright
\begin{minipage}{0.3\linewidth}
  \centerline{\includegraphics[scale=0.55]{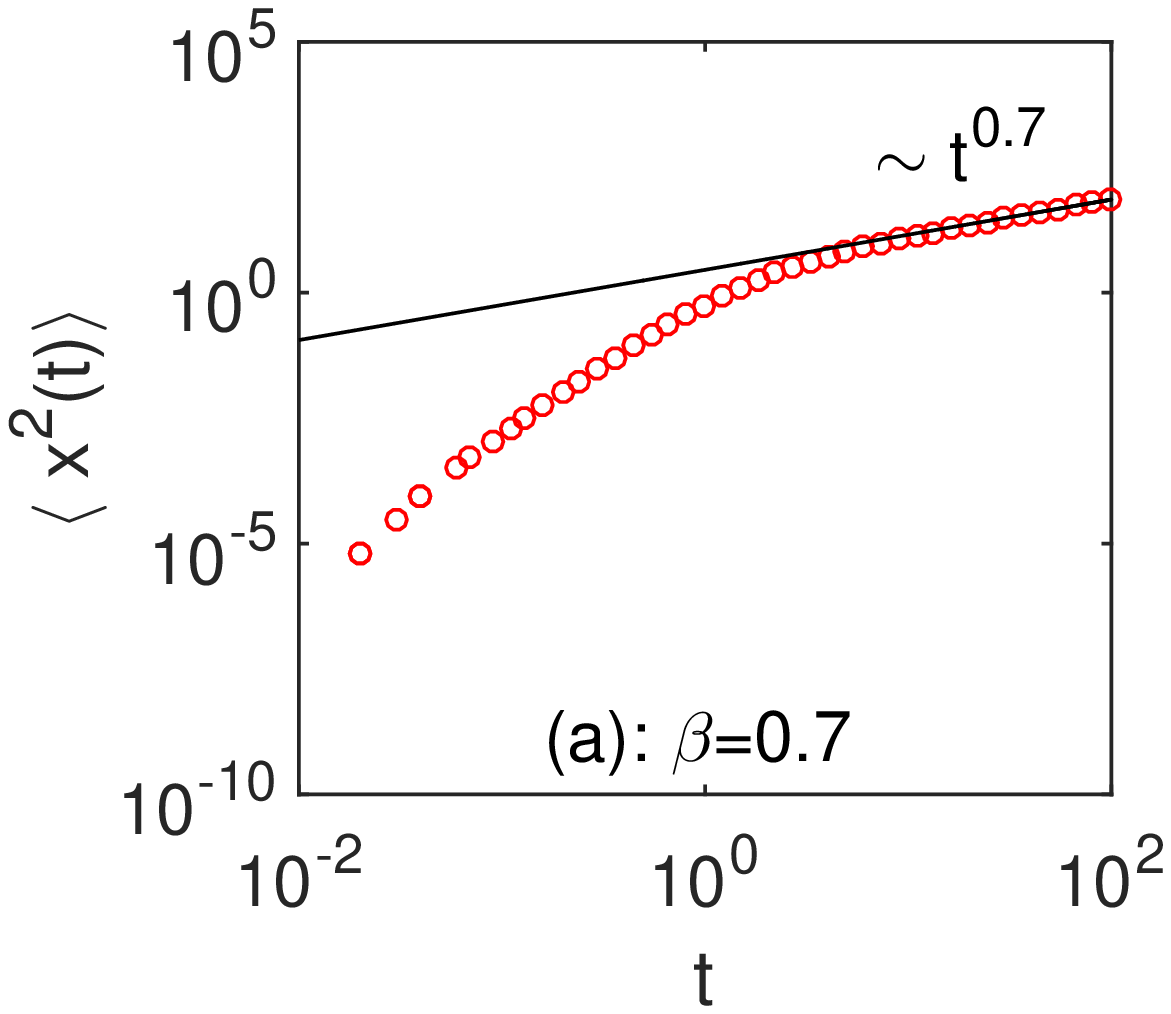}}
\end{minipage}
\hspace{8mm}
\begin{minipage}{0.4\linewidth}
  \centerline{\includegraphics[scale=0.55]{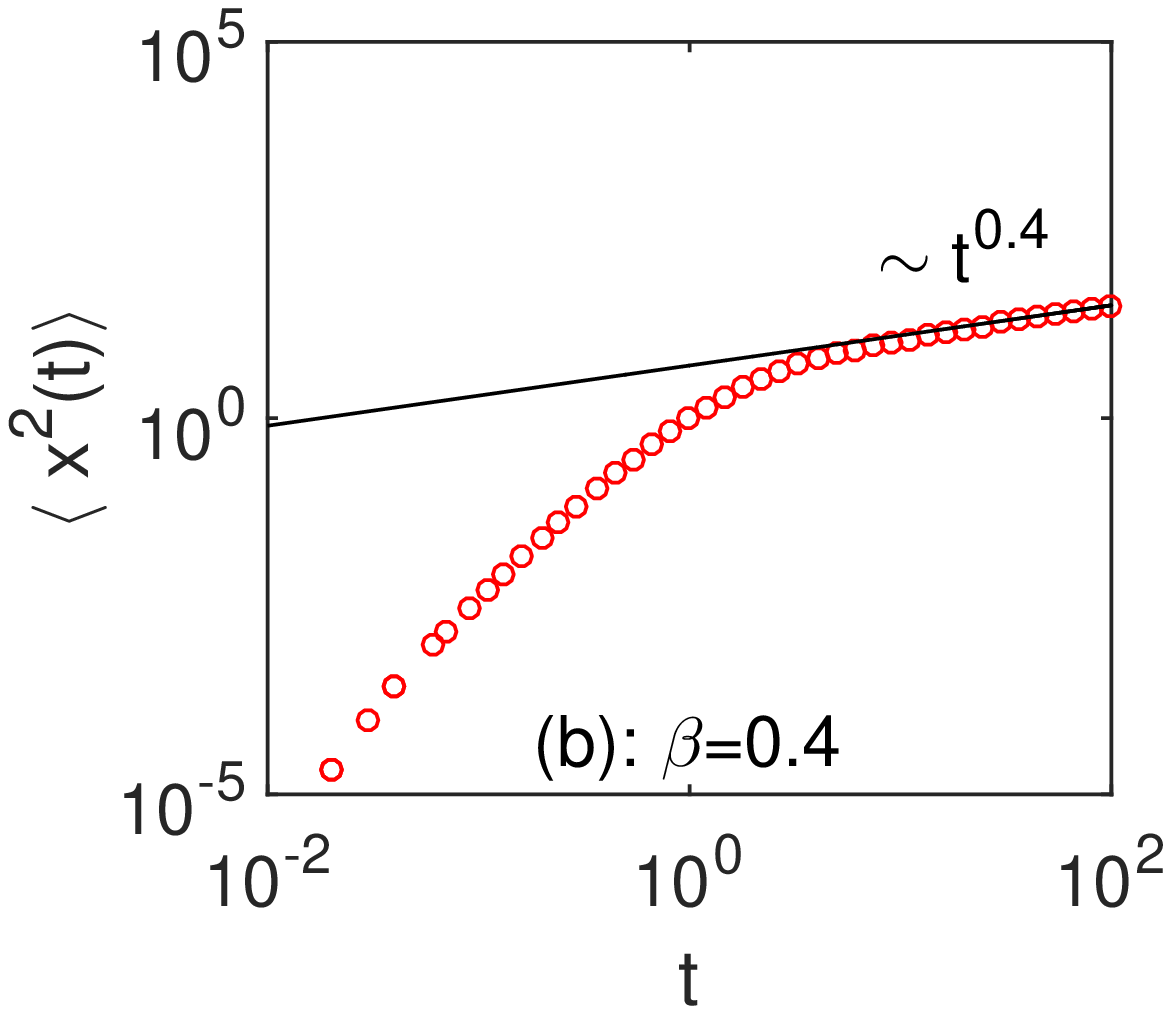}}
\end{minipage}
\caption{Simulation results of the EAMSD for stochastic process described by the Langevin equation (\ref{langevin}) with fixed parameter $\tau=1$. Other parameters: $v_0=1$, $\nu=1$.
Red circle-markers and black solid lines represent the simulation results averaging over $10^4$ trajectories and the theoretical results (\ref{contion}), respectively.}\label{tu1}
\end{figure}

In this section, we fix constant initial velocity $v_0$ and compare the two cases with fixed and random parameters $\tau$ by investigating the diffusion behavior and ergodicity of the Langevin equation (\ref{langevin}). Let us firstly consider the constant relaxation timescale $\tau$ and initial velocity $v_0$. The velocity correlation function, as well as the variance of the velocity process, can be evaluated directly based on (\ref{covar}) for large $t_1$, i.e.,
\begin{eqnarray}\label{1}
\langle v(t_1)v(t_2)\rangle\simeq \nu\tau t_1^{\beta-1} {\rm e}^{-\frac{t_2-t_1}{\tau}},\nonumber\\
    \langle v^2(t)\rangle\simeq \nu\tau t^{\beta-1},
\end{eqnarray}
where the asymptotic formula of hypergeometric function $_1F_1(a, b, z)\simeq \frac{\Gamma(b)}{\Gamma(a)}{\rm e}^zz^{a-b}$ for large $z$ has been used.
Fixing $t_1$ and putting $t_2$ to the infinity in the first equation of (\ref{1}), the velocity correlation function decays exponentially as the Ornstein-Uhlenbeck process, although the diffusivity becomes time dependent in our model (\ref{langevin}).
For $0<\beta<1$, however, the velocity variance $\langle v^2(t)\rangle$ tends to zero at a power law rate in the second equation of (\ref{1}), while the one of the Ornstein-Uhlenbeck process converges to a constant.
Further based on the velocity correlation function, the EAMSD of the stochastic process with fixed parameter $\tau$ is
\begin{equation}\label{contion}
    \langle x^2(t)\rangle=\int_0^t\int_0^t \langle v(t_1)v(t_2)\rangle {\rm d}t_1{\rm d}t_2
    \simeq \frac{2\nu\tau^2}{\beta} t^{\beta},
\end{equation}
which is independent of the initial velocity $v_0$ in the long time asymptotics.
The stochastic process described by the Langevin equation (\ref{langevin}) with fixed parameter $\tau$ exhibits subdiffusion behavior since $0<\beta<1$. The simulation results of the EAMSD with different $\beta$ are shown in figure \ref{tu1}, which tend to the theoretical lines for large $t$.

When evaluating the ensemble-averaged TAMSD, we should additionally calculate the aging EAMSD $\langle (x(t+\Delta)-x(t))^2\rangle$ with the initial observation time $t$ which is much longer than the subsequent evolutionary time $\Delta$. More precisely, for fixed parameter $\tau$, the aging EAMSD can be obtained by integrating the velocity correlation function (\ref{covar}) as
\begin{eqnarray}\label{a}
    \langle (x(t+\Delta)-x(t))^2\rangle
    &\simeq 2\nu\tau^2t^{\beta-1}(\Delta-\tau(1-{\rm e}^{-\frac{\Delta}{\tau}}))\nonumber\\
    &\simeq\left\{
    \begin{array}{cc}
     \nu\tau  t^{\beta-1} \Delta^2, &\Delta\ll\tau,\\
     2\nu\tau^2  t^{\beta-1} \Delta, &\Delta\gg\tau,
    \end{array}\right.
\end{eqnarray}
where the detailed derivations are present in \ref{App0}.
The result in (\ref{a}) shows that the aging EAMSD exhibits the transition from ballistic diffusion at short time to normal diffusion at long time. The dependence of the initial observation time $t$ implies the aging behavior of the process $x(t)$.

\begin{figure}[!htb]
\flushright
\begin{minipage}{0.3\linewidth}
  \centerline{\includegraphics[scale=0.55]{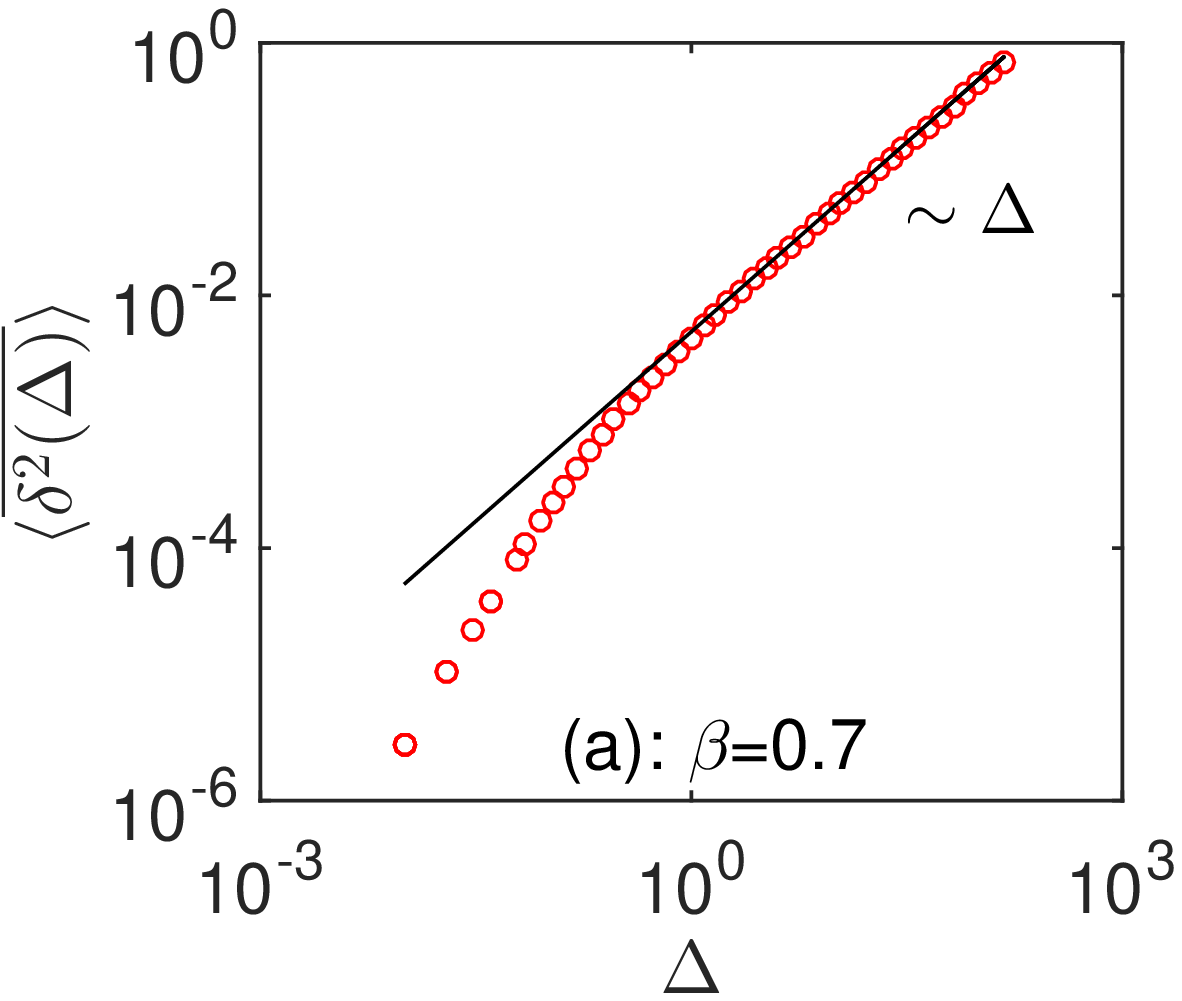}}
\end{minipage}
\hspace{8mm}
\begin{minipage}{0.4\linewidth}
  \centerline{\includegraphics[scale=0.55]{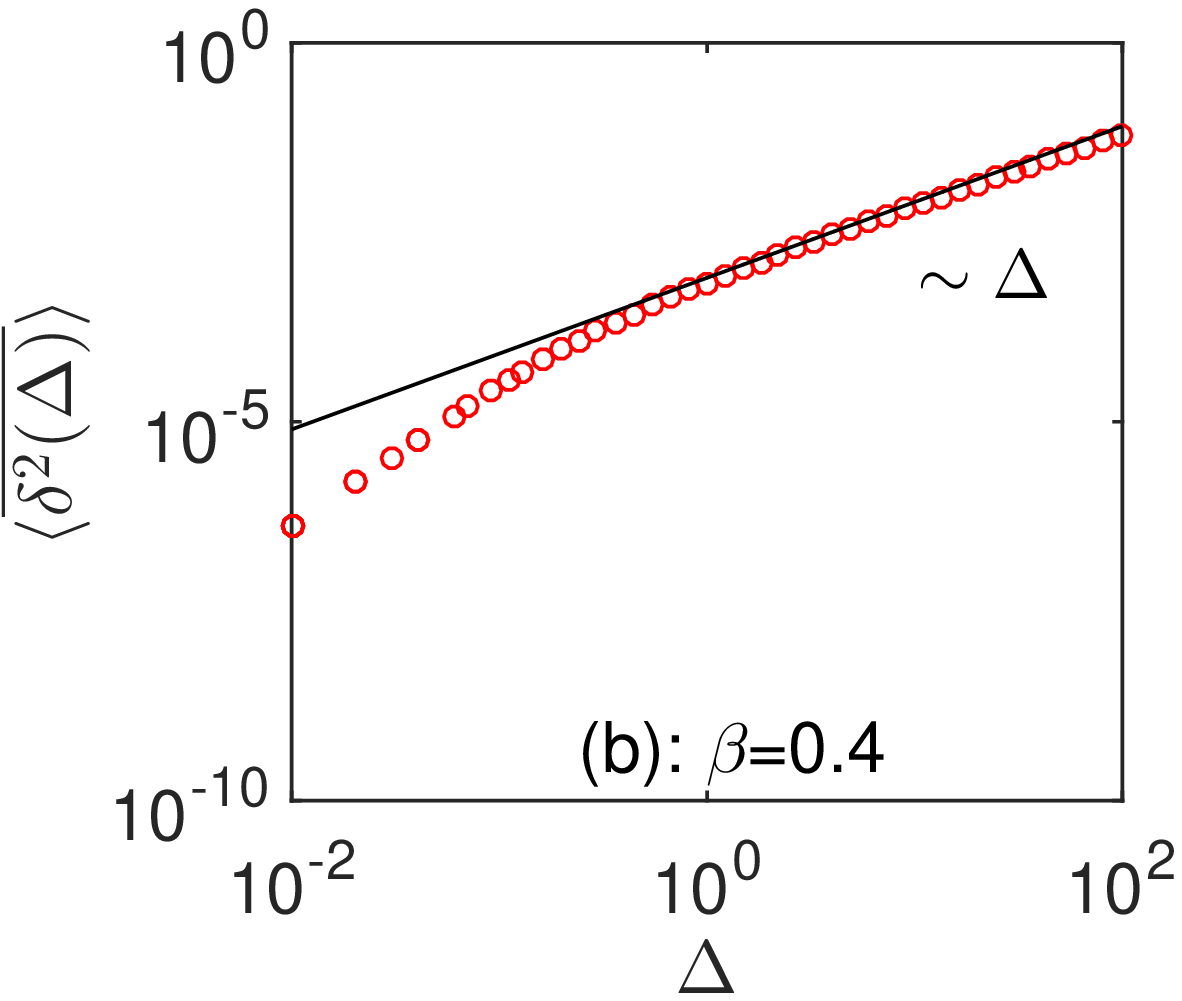}}
\end{minipage}
\caption{Simulation results of the ensemble-averaged TAMSD for the stochastic process described by the Langevin equation (\ref{langevin}) with fixed parameter $\tau=0.1$. Other parameters: $T=10^3$, $v_0=1$, $\nu=1$.
Red circle-markers and black solid lines represent the simulation results averaging over $10^3$ trajectories and the theoretical results (\ref{TAMSDTAMSD}), respectively.}\label{tu2}
\end{figure}

Based on the aging EAMSD, the ensemble-averaged TAMSD of the stochastic process for $\Delta\ll T$ can be obtained by use of the definition of TAMSD in (\ref{TADef}) as
\begin{eqnarray}\label{TAMSDTAMSD}
    \langle \overline{{\delta^2(\Delta)}}\rangle
    &=\frac{1}{T-\Delta}\int_0^{T-\Delta}  \langle( x(t+\Delta)-x(t))^2\rangle {\rm d}t\nonumber\\
    &\simeq \frac{2\nu\tau^2}{\beta}T^{\beta-1}\Delta,
\end{eqnarray}
which exhibits normal diffusion behavior with respect to the lag time $\Delta$. It shows the pronounced difference from the EAMSD in  (\ref{contion}), and thus implies the nonergodicity behavior of the Langevin equation (\ref{langevin}). Figure \ref{tu2} shows the simulation results of the ensemble-averaged TAMSDs with different $\beta$,  which coincide with the theoretical results well.

Now we turn attention to the case with random parameter $\tau$ obeying the distribution (\ref{PDF}). The velocity correlation function averaged over random parameter $\tau$ can be expressed as
\begin{equation}\label{vv0}
		\langle v(t_1)v(t_2)\rangle
		=\int_0^\infty\langle v(t_1)v(t_2)|v_0, \tau\rangle g(\tau){\rm d}\tau.
\end{equation}
Using the residue theorem (with detailed derivations in \ref{App1}), we obtain the asymptotic behavior of velocity correlation function for large $t_1$, $t_2$ and $t_1<t_2$:
\begin{eqnarray}\label{ve}
    &\langle v(t_1)v(t_2)\rangle
    \simeq C_1v_0^2 (t_1+t_2)^{-\alpha-1}  \nonumber\\
   &~~~~~~~~~~~~~~~~~~+ \frac{2 C_1\nu t_1^{\beta}(t_1+t_2)^{-\alpha-1}}{\beta} {}_2F_1\left(\alpha+1,\beta,\beta+1,\frac{2t_1}{t_1+t_2}\right),
\end{eqnarray}
where $C_1=\frac{\Gamma(1+\alpha)}{\Gamma(1-\alpha)}\frac{\langle \tau\rangle^{1+\alpha}\Gamma(1/\alpha)^\alpha}{\alpha^{\alpha-1}}$.
The Gaussian hypergeometric function is defined as
\begin{equation}
_2F_1(a,b,c,z)=\sum_{n=0}^\infty \frac{(a)_n(b)_n}{(c)_n} \frac{z^n}{n!},
\end{equation}
where $(q)_n$ is the (rising) Pochhammer symbol, and the asymptotic expression for small $z$ is: $_2F_1(a,b,c,z)\simeq 1+\frac{ab}{c}z$. Therefore, for fixed time $t_1$ and $t_2\rightarrow\infty$, the velocity correlation function in (\ref{ve}) scales as
\begin{equation}\label{q}
    \langle v(t_1)v(t_2)\rangle\simeq \frac{2\nu C_1 t_1^{\beta}}{\beta} t_2^{-\alpha-1}.
\end{equation}
By observing the velocity correlation function (\ref{q}), one finds that the correlation of velocity process at two different  times decays at a power law rate, presenting the significant difference from the exponential decaying rate in the case with fixed parameter $\tau$ and time-dependent diffusivity in (\ref{1}). The decay rate is controlled by the L\'{e}vy exponent $\alpha$ in the distribution $g(\tau)$ in (\ref{PDF}), while the coefficient is effected by the time-dependent diffusivity with power $\beta$.

Based on the velocity correlation function (\ref{ve}), the EAMSD can be obtained through double integrals as (\ref{contion}) shows:
\begin{equation}\label{E}
    \langle x^2(t)\rangle\simeq  \frac{2(1-2^{-\alpha})C_1v_0^2}{\alpha(1-\alpha)}t^{1-\alpha}+\frac{2^{2-\beta}\nu C_1 C_2}{1-\alpha+\beta}t^{1-\alpha+\beta},
\end{equation}
where $C_2=\int_0^1 (\eta+1)^{\beta-\alpha-1}\int_0^{\frac{2\eta}{\eta+1}} z^{\beta-1}(1-z)^{-\alpha-1}{\rm d}z{\rm d}\eta$ is a finite constant. The detailed derivations are present in \ref{App2}. Note that the EAMSD (\ref{E}) for random parameter $\tau$ cannot be got by performing the emsemble average over parameter $\tau$ on the EAMSD (\ref{contion}) directly, since the latter is obtained with a series of asymptotic calculations conditioned on fixed $\tau$. More analyses on the ineffectivity of the direct ensemble average over the random variable $\tau$ will be presented in the next section.

\begin{figure}[!htb]
\flushright
\begin{minipage}{0.3\linewidth}
  \centerline{\includegraphics[scale=0.55]{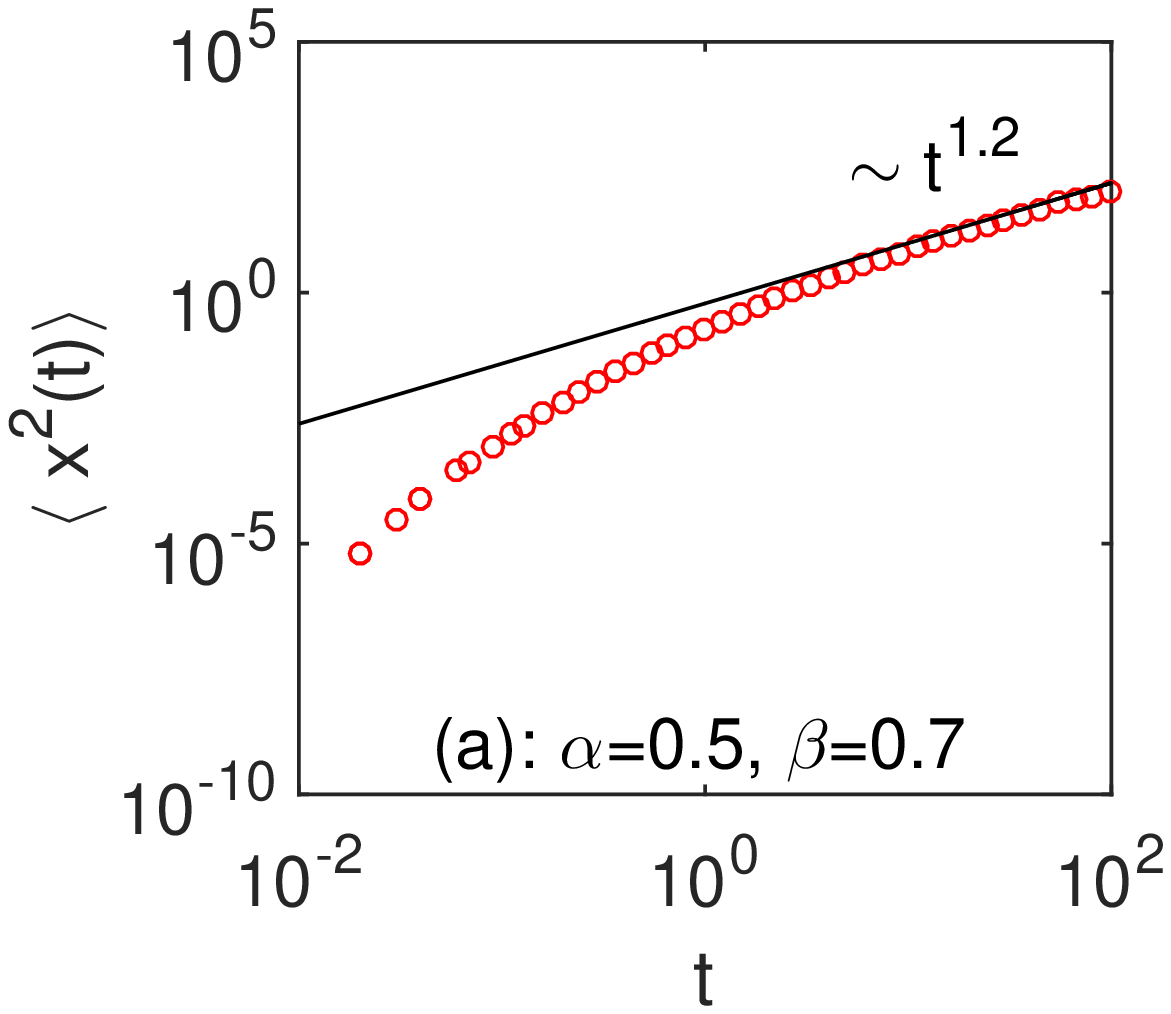}}
\end{minipage}
\hspace{9mm}
\begin{minipage}{0.4\linewidth}
  \centerline{\includegraphics[scale=0.55]{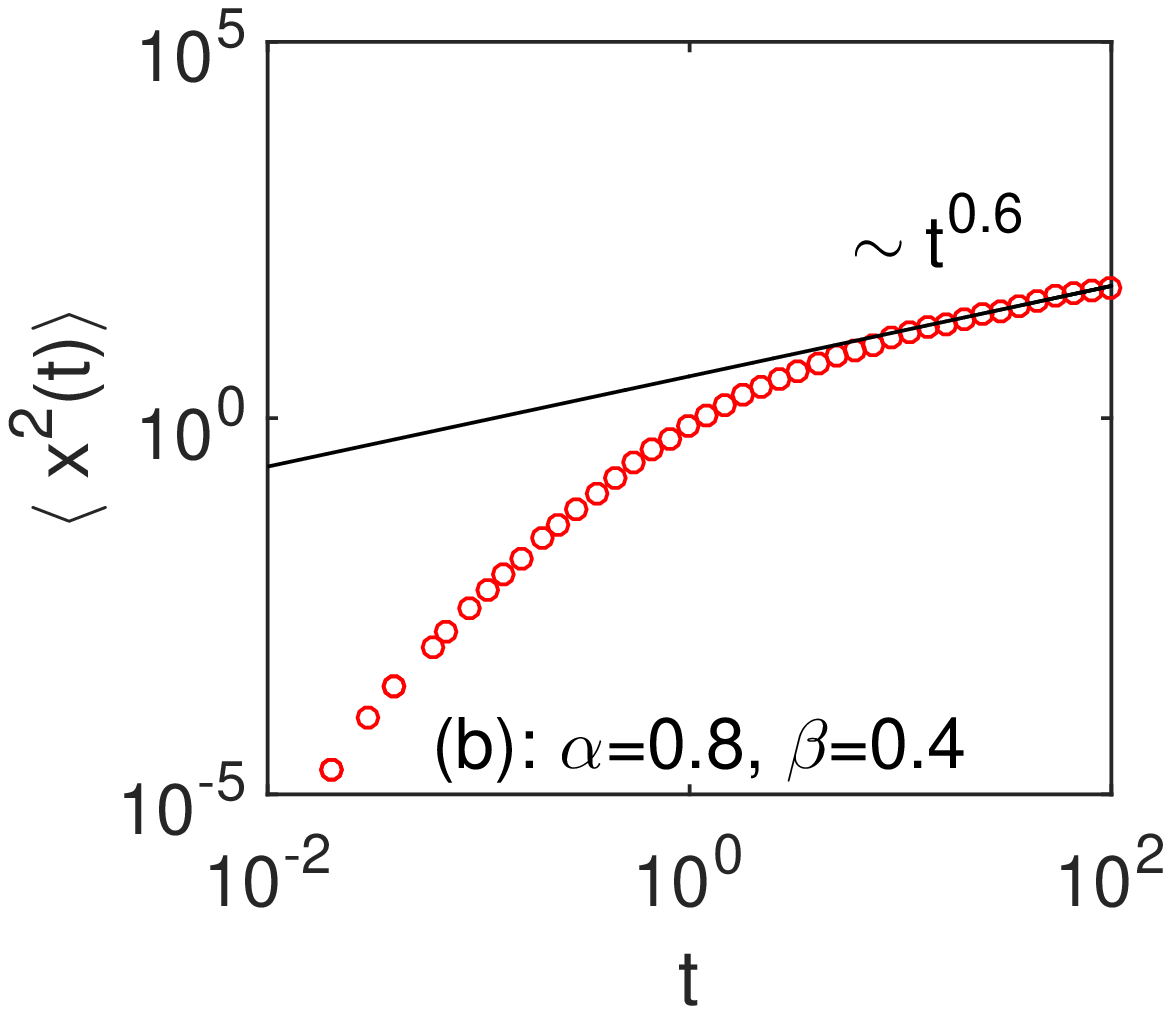}}
\end{minipage}
\begin{minipage}{0.3\linewidth}
	\centerline{\includegraphics[scale=0.55]{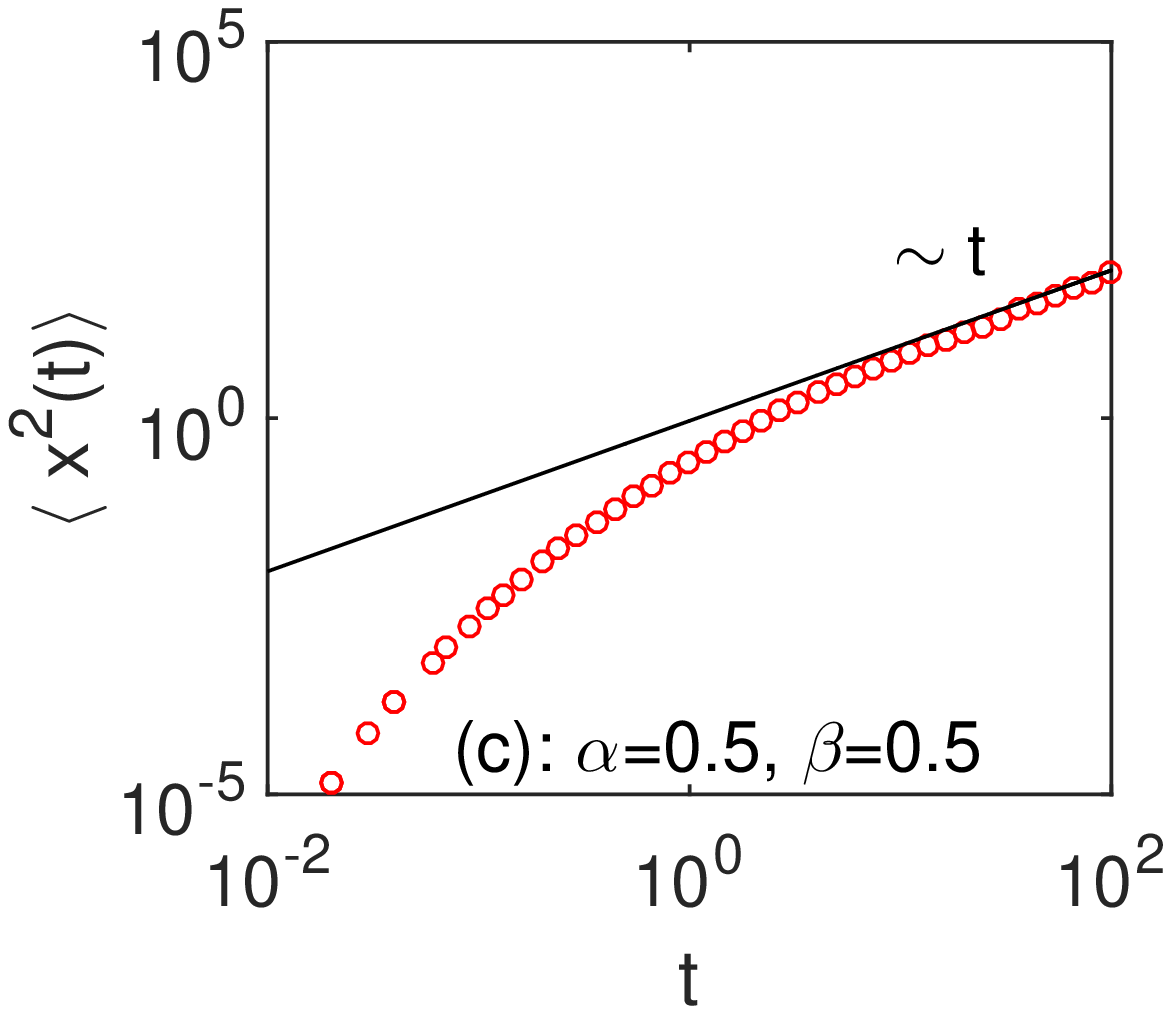}}
\end{minipage}
\hspace{9mm}
\begin{minipage}{0.4\linewidth}
	\centerline{\includegraphics[scale=0.55]{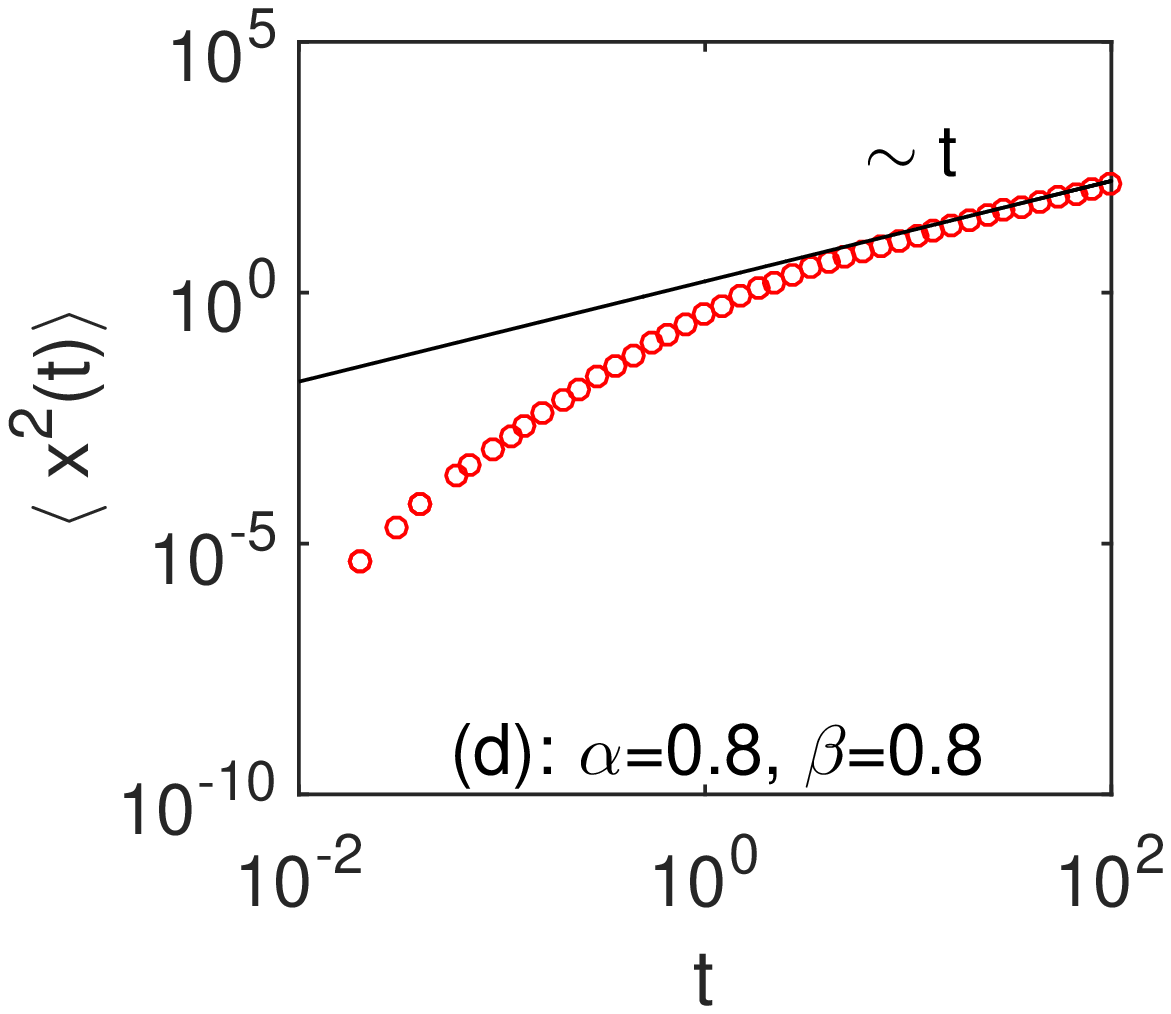}}
\end{minipage}
\caption{Simulation results of the EAMSD for stochastic process described by the Langevin equation (\ref{langevin}) with random parameter $\tau$. Other parameters: $v_0=1$, $\nu=1$, $\langle\tau\rangle=0.5$ in (a) and (c), $\langle\tau\rangle=0.9$ in (b) and (d).
Red circle-markers representing the simulation results obtained by averaging over $10^4$ trajectories, coincide with the theoretical results (\ref{EAMSD}) represented by black solid lines.}\label{tu3}
\end{figure}

Considering the relationship between the power law exponents in two terms of (\ref{E}), which is $1-\alpha<1-\alpha+\beta$,
the second term plays the dominating role at long time and the effect of the initial velocity $v_0$ can be omitted, in other words, the EAMSD of the Langevin equation (\ref{langevin}) behaves as
\begin{equation}\label{EAMSD}
    \langle x^2(t)\rangle\simeq\frac{2^{2-\beta}\nu C_1 C_2}{1-\alpha+\beta}t^{1-\alpha+\beta}.
\end{equation}
The coincidence between the theoretical and simulated results of the EAMSD is shown in figure \ref{tu3} with different $\alpha$ and $\beta$. Since the diffusion exponent satisfies $0<1-\alpha+\beta<2$, the stochastic process described by Langevin equation (\ref{langevin}) with both random relaxation timescale $\tau$ and time-dependent diffusivity $t^{\beta-1}$ can show various types of diffusion behavior, including normal diffusion ($\alpha=\beta$), subdiffusion ($\alpha>\beta$) and superdiffusion ($\alpha<\beta$).

The Langevin equation (\ref{langevin}) recovers the Brownian motion and the EAMSD (\ref{EAMSD}) becomes normal when $\alpha=\beta=1$ .
Compared with this normal case, although the parameters $\alpha$ and $\beta$ are both in the range of $(0,1)$, they provide the opposite effects on the diffusion behavior, which can be inferred from the opposite signs in front of $\alpha$ and $\beta$ in the exponent of $t$ in (\ref{EAMSD}). More precisely, the random relaxation timescale $\tau$ with parameter $\alpha$ enhances the diffusion while the time-dependent diffusivity with parameter $\beta$ suppresses the diffusion. The promoting to the diffusion of the random parameter $\tau$ can also be found by comparing the EAMSDs between (\ref{EAMSD}) and (\ref{contion}) for the cases with random and deterministic parameter $\tau$, respectively.
Since $0<\alpha,\beta<1$, the one, which is smaller or more deviates from $1$, plays the dominating role in the diffusion behavior.

\begin{figure}[!htb]
\flushright
\begin{minipage}{0.3\linewidth}
  \centerline{\includegraphics[scale=0.525]{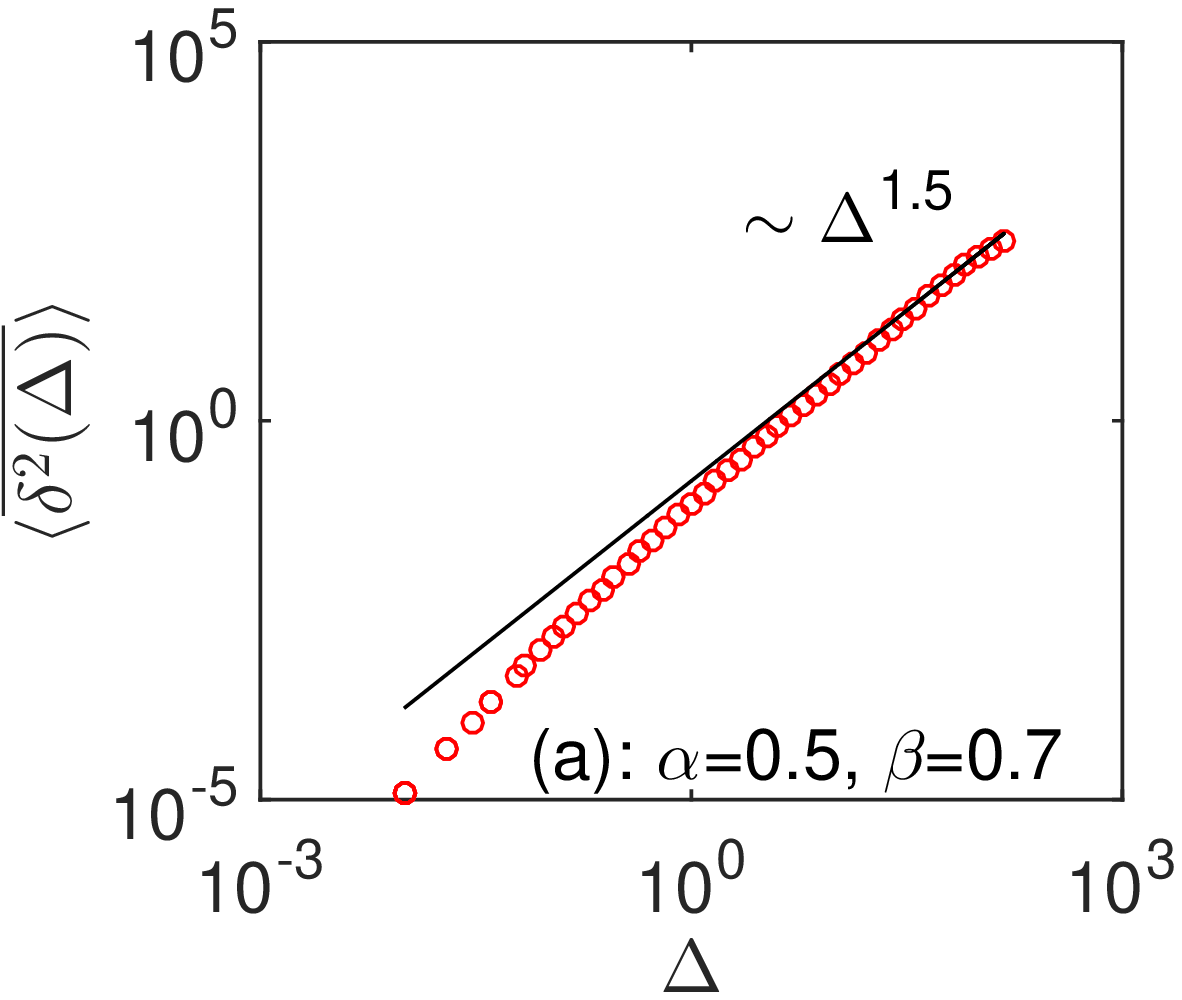}}
\end{minipage}
\hspace{7mm}
\begin{minipage}{0.4\linewidth}
  \centerline{\includegraphics[scale=0.525]{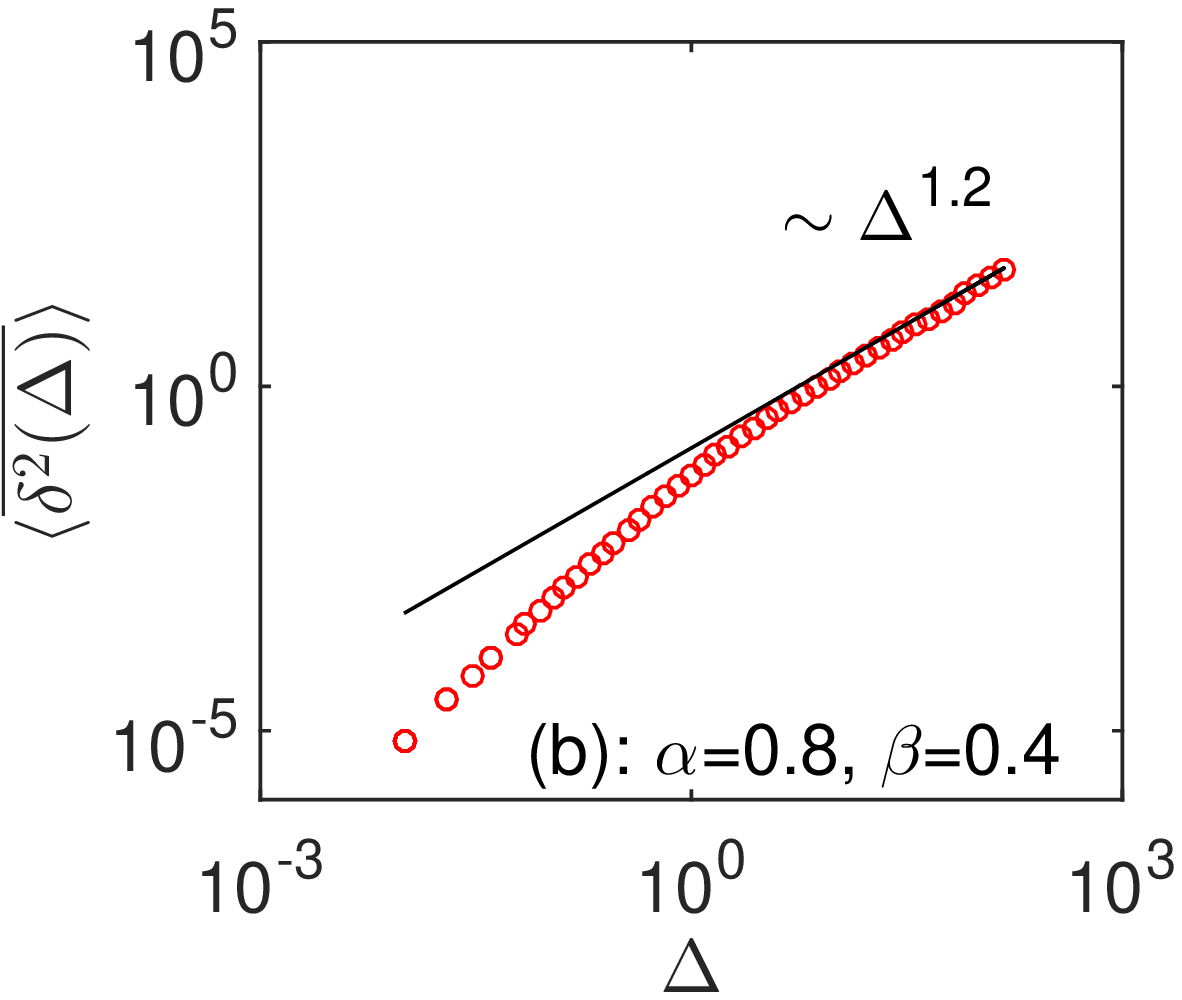}}
\end{minipage}
\caption{Simulation results of the ensemble-averaged TAMSD for stochastic process described by the Langevin equation (\ref{langevin}) with random parameter $\tau$. Other parameters: $T=10^3$, $v_0=1$, $\nu=1$, $\langle\tau\rangle=0.5$ in (a) and $\langle\tau\rangle=0.9$ in (b).
Red circle-markers, representing the simulation results obtained by averaging over $10^3$ trajectories, coincide with the theoretical results (\ref{TAMSD}) represented by black solid lines.}\label{tu4}
\end{figure}

Based on the velocity correlation function $\langle v(t_1)v(t_2)\rangle$ (\ref{ve}), similar to (\ref{a}), the aging EAMSD is, for $\Delta\ll t$,
\begin{equation}\label{aging}
    \langle (x(t+\Delta)-x(t))^2\rangle\simeq\frac{2\nu C_1}{(2-\alpha)(1-\alpha)\alpha}t^{\beta-1}\Delta^{2-\alpha},
\end{equation}
which depends on the time $t$ and indicates the aging behavior of the process $x(t)$. The interesting thing is that the parameter $\alpha$ of random variable $\tau$ only acts on the lag time $\Delta$ and does not changing the aging behavior.
The aging EAMSD (\ref{aging}) shows the superdiffusion behavior, which is faster than the case with fixed parameter $\tau$ showing normal diffusion (\ref{a}). Substituting the aging EAMSD into the definition of TAMSD and calculating the integral, we arrive at the ensemble-averaged TAMSD
\begin{equation}\label{TAMSD}
    \langle \overline{{\delta^2(\Delta)}}\rangle\simeq
    \frac{2\nu C_1}{(2-\alpha)(1-\alpha)\alpha\beta}T^{\beta-1}\Delta^{2-\alpha}.
\end{equation}
The corresponding simulation results are presented in figure \ref{tu4}. Comparing the ensemble-averaged TAMSDs in (\ref{TAMSDTAMSD}) and (\ref{TAMSD}), one can see that the randomness of parameter $\tau$ accelerates the diffusion behavior in the sense of TAMSD (from order $1$ to order $2-\alpha$). The inequality of the EAMSD (\ref{EAMSD}) and TAMSD (\ref{TAMSD}) means the ergodicity breaking of the Langevin equation (\ref{langevin}) with respect to the MSD.

\section{Fixed and random initial velocity $v_0$}\label{Sec3}
The results in section \ref{Sec1} are obtained in the condition of a fixed initial velocity $v_0$, where the two terms in the EAMSD (\ref{E}) correspond to those in the velocity correlation function $\langle v(t_1)v(t_2)\rangle$ (\ref{ve}), respectively, one containing $v_0^2$ and another one not. The first terms in (\ref{ve}) and (\ref{E}) can be neglected compared with the corresponding second terms when $v_0$ is a constant. Actually, for the case with random initial velocity $v_0$, as long as it is independent of the relaxation timescale $\tau$ and has a finite second moment, the first terms in (\ref{ve}) and (\ref{E}) can be omitted compared with the second terms after performing the ensemble average with respect to $v_0$.

On the other hand, if the initial velocity $v_0$ is not only random, but also depends on relaxation timescale $\tau$, then the ensemble average with respect to $\tau$ on (\ref{vv0}) will also work on $v_0^2$ and the result will be different.
Let us consider a special case $v_0^2=\nu\tau$, which implies that the two random variable $v_0$ and $\tau$ are closely related and $v_0$ has a finite second moment. For the particle's motion described by Langevin equation with random $v_0$ and $\tau$ satisfying $v_0^2=\nu\tau$, each particle has its own fixed initial velocity $v_0$ and relaxation timescale $\tau$ throughout its whole trajectory, and the pair of parameters ($v_0,\tau$) satisfy the quantitative relation $v_0^2=\nu\tau$.  But different particles have discrepant parameter pair ($v_0,\tau$).

Before performing the calculations of EAMSD and TAMSD for the case of $v_0^2=\nu\tau$, let us see the reason of choosing such a $v_0$.
Taking $\beta=1$ in (\ref{covar}) leads to the conditional velocity correlation function of the Langevin system with time-dependent diffusivity:
\begin{equation}\label{Cor2}
     \langle v(t_1)v(t_2)|v_0,\tau\rangle=(v_0^2-\nu\tau){\rm e}^{-\frac{t_1+t_2}{\tau}}
     +\nu\tau {\rm e}^{-\frac{|t_1-t_2|}{\tau}},
\end{equation}
where the average is only made over the Gaussian white noise $\xi(t)$. For fixed $v_0$ and $\tau$, the second term on the right-hand side dominates the velocity correlation function at long time, that is, the velocity correlation function is asymptotic stationary. When $\tau$ is random, however, the distribution of $\tau$ in (\ref{PDF}) implies that $\tau$ can take an arbitrary large value, and the first term cannot be ignored even for large $t_1$ and $t_2$. But when $v_0^2=\nu\tau$, the first term vanishes and the conditional velocity correlation function is stationary from the beginning of the evolution whether $\tau$ is fixed or random. Although we consider $0<\beta<1$ in this paper, the first term containing $v_0^2$ in velocity correlation function (\ref{vv0}) is independent of parameter $\beta$.
That is why we still choose the initial condition satisfying $v_0^2=\nu\tau$.

In the case of $v_0^2=\nu\tau$ for any $0<\beta<1$, when performing the average over random parameter $\tau$ in (\ref{vv0}), the first term containing $v_0^2$ in (\ref{ve}) should be modified as
\begin{eqnarray}\label{vv1}
		\int_0^\infty v_0^2{\rm e}^{-\frac{t_1+t_2}{2}} g(\tau){\rm d}\tau
		=\int_0^\infty \nu\tau {\rm e}^{-\frac{t_1+t_2}{2}} g(\tau){\rm d}\tau
		=\frac{C_1\nu}{\alpha}(t_1+t_2)^{-\alpha},
\end{eqnarray}
which is obtained by using residue theorem with the similar procedures in \ref{App1}.
When evaluating the EAMSD by performing the double integrals over velocity correlation function  $\langle v(t_1)v(t_2)\rangle$, the term in (\ref{vv1}) contributes to
\begin{equation}\label{EAMSD0}
    \langle x^2(t)\rangle_0
    \simeq\frac{(2^{2-\alpha}-2)C_1\nu}{(2-\alpha)(1-\alpha)\alpha}t^{2-\alpha}.
\end{equation}
Similarly, the term in (\ref{vv1}) has another contribution
\begin{equation}\label{EATAMSD0}
	\langle \overline{{\delta^2(\Delta)}}\rangle_0
	\simeq \frac{2^{-\alpha}C_1\nu}{(1-\alpha)\alpha}\frac{\Delta^2}{T^\alpha}
\end{equation}
to the ensemble-averaged TAMSD. Here, we use the subscript ``$0$'' to denote the part coming from random initial velocity $v_0$.

\begin{figure}[!htb]
\flushright
	\begin{minipage}{0.3\linewidth}
		\centerline{\includegraphics[scale=0.4]{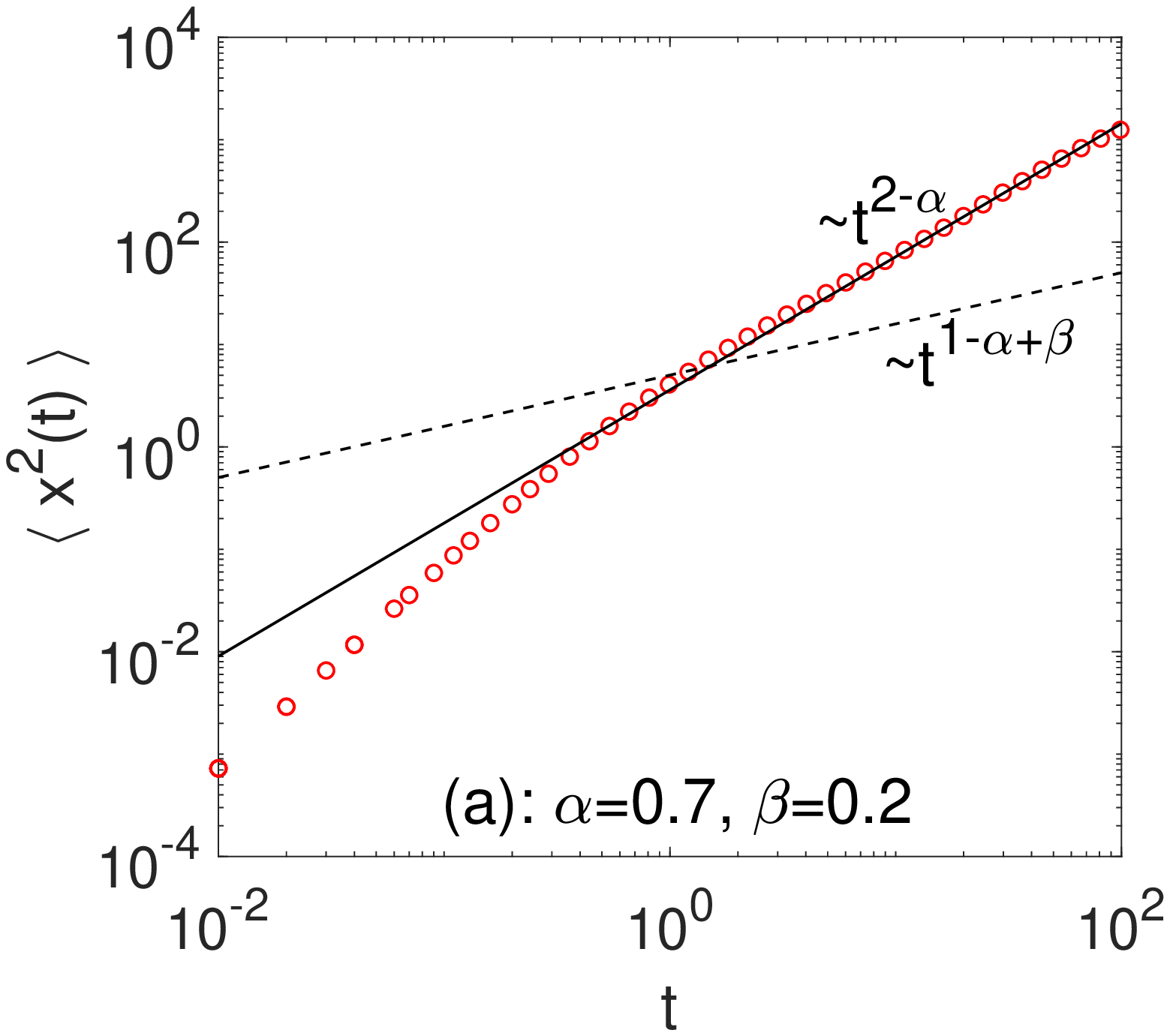}}
	\end{minipage}
	\hspace{9mm}
	\begin{minipage}{0.4\linewidth}
		\centerline{\includegraphics[scale=0.4]{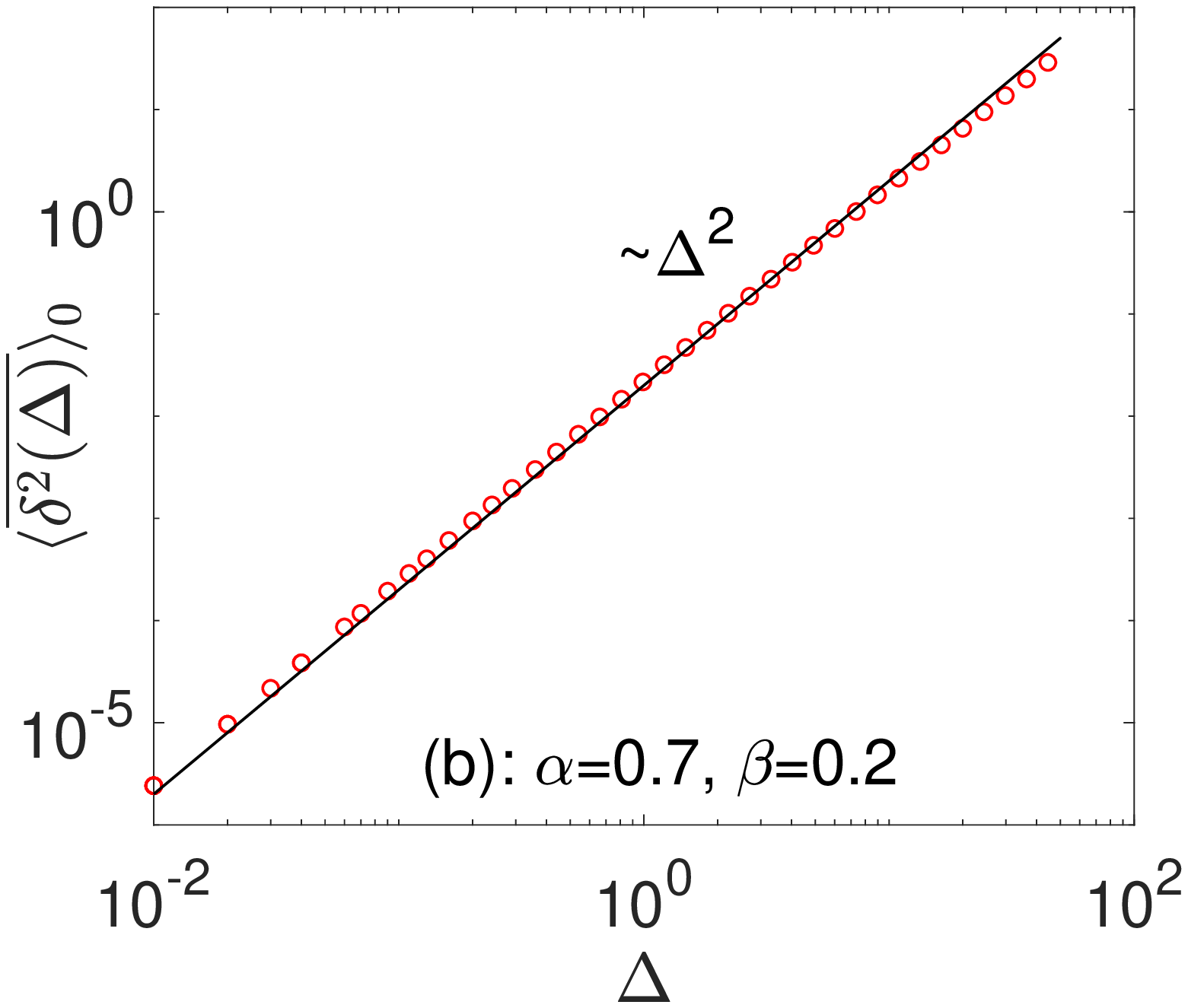}}
	\end{minipage}
	\caption{Simulation results of the total EAMSD and the partial ensemble-averaged TAMSD for stochastic process described by the Langevin equation (\ref{langevin}) with random parameters $\tau$ and $v_0$. The initial velocity $v_0$ are chosen according to the relaxation timescale $\tau$ with the relationship $v_0^2=\nu\tau$ for each trajectory. Both $\alpha=0.7$ and $\beta=0.2$ in (a) and (b).
	Other parameters: $t=T=10^2$, $\nu=1$, and $\langle\tau\rangle=0.8$.
	Red circle-markers are the simulation results averaging over $10^4$ trajectories. The black lines in left and right panels denote the theoretical results (\ref{EAMSDFinal}) and (\ref{EATAMSD0}), respectively. The black dashed line in left panel represents the subleading term of EAMSD in (\ref{EAMSD}).}\label{tu6}
\end{figure}

Comparing the EAMSDs between (\ref{EAMSD}) and (\ref{EAMSD0}), we find that the latter contributed by the random initial velocity $v_0$ plays the dominating role, i.e., the total EAMSD is
\begin{equation}\label{EAMSDFinal}
		\langle x^2(t)\rangle	\simeq\frac{(2^{2-\alpha}-2)C_1\nu}{(2-\alpha)(1-\alpha)\alpha}t^{2-\alpha}.
\end{equation}
However, the magnitude relation between ensemble-averaged TAMSDs in (\ref{TAMSD}) and (\ref{EATAMSD0}) cannot be easily determined. To verify the correctness of the partial ensemble-averaged TAMSD (\ref{EATAMSD0}) contributed by the random initial velocity, we make the simulations on the quantity
\begin{equation}\label{EATAMSD01}
	\langle \overline{{\delta^2(\Delta)}}\rangle_0
\simeq 	\langle \overline{{\delta^2(\Delta)}}\rangle
-\frac{2\nu C_1}{(2-\alpha)(1-\alpha)\alpha\beta}T^{\beta-1}\Delta^{2-\alpha},
\end{equation}
where the first term on the right-hand side is the real ensemble-averaged TAMSD and the second term is the part coming from Gaussian white noise obtained in previous section as (\ref{TAMSD}) shows. The corresponding simulations for EAMSD and ensemble-averaged TAMSD are present in figure \ref{tu6}, which are consistent to the theoretical results (\ref{EAMSDFinal}) and (\ref{EATAMSD0}), respectively.

The ensemble-averaged TAMSD becomes simple for the case with time-independent diffusivity ($\beta=1$). In this case, both the EAMSDs in (\ref{EAMSD}) and (\ref{EAMSD0}) behaves as $t^{2-\alpha}$, and the random initial velocity $v_0$ does not change the diffusion behavior. As to the ensemble-averaged TAMSD, the second term $\Delta^{2-\alpha}$ on the right-hand side of (\ref{EATAMSD01}) plays the leading role compared with the term $\Delta^2T^{\alpha}$ in $\langle \overline{{\delta^2(\Delta)}}\rangle_0$.
The difference between the EAMSD (\ref{EAMSDFinal}) and the ensemble-averaged TAMSD (\ref{EATAMSD0}) or (\ref{TAMSD}) implies the nonergodicity of the model (\ref{langevin}) with random
initial velocity $v_0$ and random relaxation timescale $\tau$. However, the special case $\beta=1$ may lead to a misunderstanding, where the EAMSD and the ensemble-averaged TAMSD both behave as
\begin{equation}\label{EA-TA-beta1}
	\langle x^2(\Delta)\rangle\simeq \langle \overline{{\delta^2(\Delta)}}\rangle\simeq
	\frac{2\nu C_1}{(2-\alpha)(1-\alpha)\alpha}\Delta^{2-\alpha}.
\end{equation}

\begin{figure}[!htb]
\centering
\includegraphics[scale=0.56]{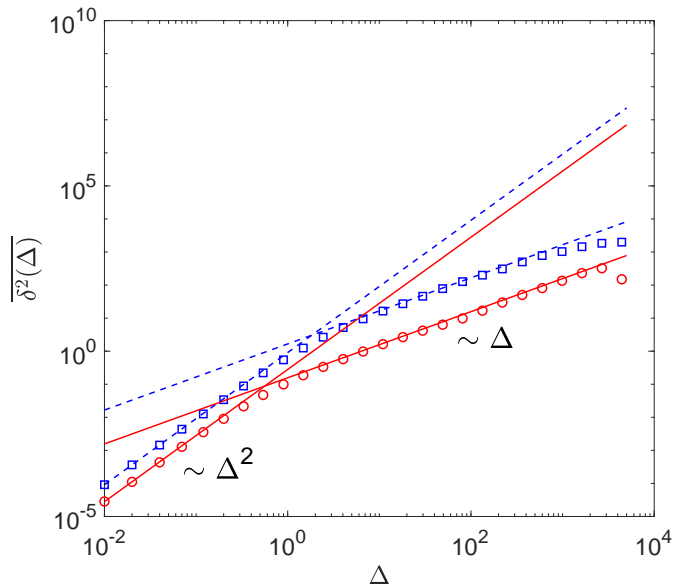}
\caption{Simulation results of the TAMSD for two trajectories of the Langevin equation (\ref{langevin}) with random parameter $\tau$. The parameters are: $\alpha=0.7$, $T=10^4$, $v_0=1$, $\nu=1$, and $\langle\tau\rangle=0.8$.
Red circle-markers and blue squared-markers denote the simulated TAMSDs $\overline{\delta^2(\Delta)}$ of two trajectories. The red solid lines and blue dashed lines represent the theoretical results of TAMSD at short time and long time in (\ref{TAMSDBM}). The simulated results agree with theoretical results very well. Since the last several markers denote large $\Delta$, which does not satisfy $\Delta\ll T$, they are not consistent to the theoretical lines. It implies the nonergodicity of the Langevin equation (\ref{langevin}) since the two trajectories are not coincident.}\label{tu7}
\end{figure}

It seems that the Langevin equation (\ref{langevin}) is ergodic due to the equivalence between the EAMSD and the ensemble-averaged TAMSD as (\ref{EA-TA-beta1}) shows. However, the TAMSD is usually a random variable for anomalous diffusion processes. To detect the ergodicity accurately, we should evaluate the TAMSD to judge whether the TAMSD converges to its ensemble average as the measurement time $T\rightarrow\infty$. For the Langevin equation (\ref{langevin}) with random parameter $\tau$, each particle (or trajectory) has a fixed $\tau$, namely, $\tau_i$ for the $i$-th particle. Therefore, each particle undergoes the Brownian motion with a fixed relaxation timescale $\tau_i$. But different particles own different parameter $\tau_i$, the specific value of which is yielded from the distribution $g(\tau)$ in (\ref{PDF}).
Further due to the ergodicity of Brownian motion, the TAMSD of each particle converges to the ensemble-averaged TAMSD with fixed relaxation timescale $\tau_i$, i.e.,
\begin{eqnarray}\label{TAMSDBM}
	\overline{\delta_i^2(\Delta)}
	    &\simeq\left\{
	\begin{array}{cc}
		\nu\tau_i  \Delta^2, &~\Delta\ll\tau_i,\\
		2\nu\tau_i^2  \Delta, &~\Delta\gg\tau_i,
	\end{array}\right.
\end{eqnarray}
which is the classical result of Brownian motion \cite{CoffeyKalmykovWaldron:2004}. It can  also be obtained from the aging EAMSD (\ref{a}) by taking $\beta=1$ and performing the integral with respect to $t$.
We make the sample of two trajectories with $\tau_1=0.22$ and $\tau_2=0.57$, and present the corresponding TAMSDs in figure \ref{tu7}, which agree with the theoretical lines very well.
For both short time and long time, the TAMSD of two trajectories are
parallel, since the difference between them is just the relaxation timescale $\tau_i$. The inconsistence of two trajectories implies the nonergodicity of the Langevin equation (\ref{langevin}) when $\beta=1$.

When we observe the ensemble-averaged TAMSD for large $\Delta$, it seems strange that the result $\Delta^{2-\alpha}$ in (\ref{EA-TA-beta1}) cannot be obtained by directly performing average over the result of asymptotic TAMSD ($\Delta^2$ or $\Delta$) in (\ref{TAMSDBM}). Actually, due to the randomness of parameter $\tau$, each trajectory owns a different but fixed $\tau_i$. Therefore, given $\Delta$, the large amounts of trajectories can be divided into two categories, one with $\tau_i$ larger than $\Delta$ and another one smaller than $\Delta$. Thus, the ensemble average should not be performed solely on any one of the terms in (\ref{TAMSDBM}), but
\begin{eqnarray}\label{TAMSDDivide}
	\langle \overline{{\delta^2(\Delta)}}\rangle
	&= \int_0^\infty\overline{{\delta^2(\Delta)}}g(\tau){\rm d}\tau \nonumber\\
	&\simeq \int_0^\Delta 2\nu\tau^2\Delta g(\tau){\rm d}\tau
	   +  \int_\Delta^\infty \nu\tau\Delta^2 g(\tau){\rm d}\tau,
\end{eqnarray}
where the two integrands come from the different asymptotics in (\ref{TAMSDBM}). The two terms in (\ref{TAMSDDivide}) can both be verified to scale as $\Delta^{2-\alpha}$ as (\ref{EA-TA-beta1}) shows. The detailed calculations are presented in \ref{App3}.

\begin{figure}
  \centering
	\includegraphics[scale=0.53]{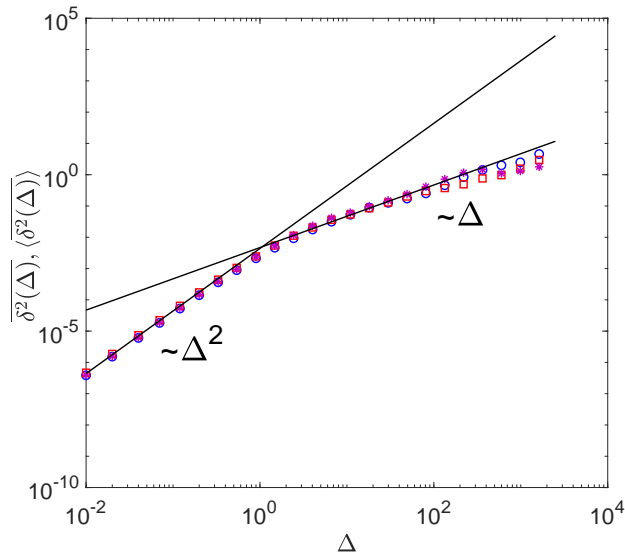}
	\caption{Simulation results of the TAMSD for three trajectories of the Langevin equation (\ref{langevin}) with the same initial velocity $v_0$ and relaxation timescale $\tau$. The parameters are: $\alpha=0.7$, $\beta=0.2$, $T=10^4$, $v_0=0$, $\nu=1$, and $\langle\tau\rangle=0.8$.
	Different markers denote the simulated TAMSDs $\overline{\delta^2(\Delta)}$ of the three trajectories, while the black lines represent the theoretical results of TAMSD at short time and long time in (\ref{TAMSDSBM}). The three TAMSD from different trajectories overlap and agree with theoretical results well.
	Several markers with large $\Delta$ deviate slightly from the theoretical line due to the loss of condition $\Delta\ll T$.
	The consistence of the TAMSD for different trajectories implies the self-average property of the Langevin equation (\ref{langevin}) with fixed $v_0$ and $\tau$.}
	\label{tu8}
\end{figure}

The analyses around (\ref{TAMSDDivide}) are also valid for the case $\beta<1$, where the TAMSD still has the good property of self-averaging. According to (\ref{a}), the corresponding TAMSD for each trajectory is
\begin{equation}\label{TAMSDSBM} \overline{\delta_i^2(\Delta)}_\beta=\frac{T^{\beta-1}}{\beta}\overline{\delta_i^2(\Delta)},
\end{equation}
which is different from the result (\ref{TAMSDBM}) by a constant. The self-averaging property of the TAMSD and its equivalence to the theoretical result (\ref{TAMSDSBM}) has been verified through the simulations as figure \ref{tu8} shows.

The aim of making different assumptions about the initial velocity $v_0$
is to detect how the initial ensemble influences the diffusion behavior. For the underdamped Langevin equation describing the classical Brownian motion with fixed $\tau$, a fixed $v_0$ is named non-equilibrated initial ensemble, which means all particles are introduced to the system at $t=0$. On the other hand, if the system reaches a certain equilibrium before we start measuring it, namely, the initial velocity $v_0$ has reached at the equilibrium distribution at $t=0$, then it is named equilibrated initial ensemble \cite{KlafterZumofen:1993}. Different initial ensemble usually leads to different diffusion behavior, which has been discussed a lot \cite{AkimotoCherstvyMetzler:2018,LeibovichBarkai:2013,HidalgoBarkaiBurov:2021,FroembergSchmiedebergBarkaiZaburdaev:2015,WangChenDeng:2019,WangChenDeng:2019-2,HouCherstvyMetzlerAkimoto:2018}.
It is noteworthy that the way of initial ensemble affecting the diffusion behavior is quite special for the underdamped Langevin equation (\ref{langevin}) with random relaxation timescale $\tau$.

\section{Summary}\label{Sec4}

The rich dynamics of particle's motion in biological systems have been attracting the attention of many researchers in the field of statistical physics. The complexity of the cell environment brings many difficulties to the research of building models to describe the particle's motion and further theoretical analyses.
In this paper, the particle's motion is characterized by an underdamped Langevin equation with a time-dependent diffusivity of velocity, in which the random relaxation timescale $\tau$ describes the complexity of the medium, and the random initial velocity $v_0$ represents a special initial ensemble.
The aim of this paper is to detect how the two parameters, relaxation timescale $\tau$ and initial velocity $v_0$, influence the diffusion behavior and ergodic property of the Langevin system.

For the relaxation timescale $\tau$, we find its randomness
slows down the decay rate of the velocity correlation function, and enhances the diffusion behavior of the Langevin system (\ref{langevin}), which is opposite to the effect of the time-dependent diffusivity $t^{\beta-1},~0<\beta<1$. Therefore, the exponents $\alpha$ and $\beta$ play the competition roles and the stochastic process can present different diffusion behavior (from subdiffusion to superdiffusion) with different values of $\alpha$ and $\beta$.
The smaller one between $\alpha$ and $\beta$ plays the dominating role in the diffusion process.
As to the ergodic property, it is the time-dependent diffusivity $t^{\beta-1}$, rather than the random relaxation timescale $\tau$, who leads to the aging behavior. Therefore, the Langevin system is nonergodic whether $\tau$ is fixed or random.

For the initial velocity $v_0$, its randomness brings a very novel phenomenon. We mainly study the special case $v_0^2=\nu\tau$, which implies that the two random variable $v_0$ and $\tau$ are closely related. The random $v_0$ enhances the diffusion behavior in (\ref{EAMSD0}) (from $t^{1-\alpha+\beta}$ to $t^{2-\alpha}$) with respect to EAMSD. It also contributes to a ballistic term $\Delta^2T^{-\alpha}$ (\ref{EATAMSD0}) to the ensemble-averaged TAMSD which cannot be omitted even for long time.
Further, we investigate the special case $\beta=1$, i.e., the diffusivity of velocity becomes a constant. Surprisingly, the random initial velocity $v_0$ does not change the diffusion any more when $\beta=1$.

The random parameters in physical models are sometimes underestimated both from the aspects of the calculation procedure and the dynamic behavior. For the former aspect, we point that the quantity cannot be evaluated by directly performing ensemble average over the random parameters on the original quantity, and analyse the reasons around (\ref{TAMSDDivide}). For the latter aspect, although the relaxation timescale $\tau$ of the Langevin system (\ref{langevin}) has finite first moment and the initial velocity $v_0$ has finite second moment, both of them change the diffusion behavior of particles and bring us the novel phenomena.
The detailed derivations and meticulous analyses in this paper enormously enrich the theoretical part of physical models with random parameters.

\section*{Acknowledgments}
This work was supported by the National Natural Science
Foundation of China under Grant No. 12105145, the Natural Science Foundation of Jiangsu Province under Grant No. BK20210325, and
the Fundamental Research Funds for the Central Universities under Grants No. lzujbky-2020-it02.

\appendix	
\section{Normalization and mean of random variable $\tau$ obeying distribution (\ref{PDF}), together with the corresponding algorithm}\label{App-1}

Let us firstly see the mean value of the random relaxation timescale $\tau$ obeying the distribution (\ref{PDF}). For convenience, taking $C=\frac{\alpha}{\Gamma(1/\alpha)\langle\tau\rangle}$, we have
\begin{equation}
	g(\tau)=\frac{C\langle\tau\rangle}{\tau}L_\alpha(C\tau).
\end{equation}
Thus, the mean value of $\tau$ is
\begin{eqnarray}
\int_0^\infty \tau g(\tau){\rm d}\tau&=\int_0^\infty C\langle\tau\rangle L_\alpha(C\tau){\rm d}\tau
=\langle\tau\rangle \int_0^\infty L_\alpha(\tau){\rm d}\tau
=\langle\tau\rangle,
\end{eqnarray}
where we have used the variable substitution in the second equality and the normalization of L\'{e}vy distribution in the last equality. Then we turn to the normalization of the random variable $\tau$, where some technical procedures will be performed. The Laplace transform of $g(\tau)$ is
\begin{eqnarray}
		\hat{g}(\lambda)&=\int_0^\infty {\rm e}^{-\lambda\tau}g(\tau){\rm d}\tau \nonumber\\
		&=C\langle\tau\rangle \int_0^\infty\int_\lambda^\infty {\rm e}^{-\mu\tau}{\rm d}\mu L_\alpha(C\tau){\rm d}\tau \nonumber\\
		&=C\langle\tau\rangle \int_\lambda^\infty {\rm d}\mu \int_0^\infty  {\rm e}^{-\mu\tau} L_\alpha(C\tau){\rm d}\tau \nonumber\\
		&=\langle\tau\rangle \int_\lambda^\infty {\rm e}^{-(\mu/C)^\alpha}{\rm d}\mu,
\end{eqnarray}
where we have used the formula ${\rm e}^{-\lambda\tau}/\tau=\int_\lambda^\infty {\rm e}^{-\mu\tau}{\rm d}\mu$ in the second line, changed the order of integrating in the third line and performed the Laplace transform of L\'{e}vy distribution in the last line. The normalization of random variable $\tau$ can be verified by taking $\lambda=0$:
\begin{equation}
	\hat{g}(0)=\langle\tau\rangle \int_0^\infty {\rm e}^{-(\mu/C)^\alpha}{\rm d}\mu=1.
\end{equation}

The key of simulating the trajectories of Langevin equation (\ref{langevin}) is to generate the random variable $\tau$ obeying distribution $g(\tau)$. The first step of generating $\tau$ is to obtain the function $L_\alpha(x)$ numerically (i.e., the values $L_\alpha(x_i)$ at every point $x_i$) by performing the inverse Laplace transform on $\rm{e}^{-\lambda^\alpha}$ through MATLAB. Then we obtain the function $g(\tau_i)$ numerically through linear transform on $L_\alpha(x_i)$ with $x_i=C\tau_i$. Further, we sum $g(\tau_i)$ to obtain the numerical cumulative distribution function $F(\tau_i)$, together with its inverse function $F^{-1}(z),~z\in[0,1]$. Finally, by use of the well-known inverse transform sampling method, we can generate the random variable $\tau$ obeying the distribution $g(\tau)$ in the way that
\begin{equation}
	\tau=F^{-1}(U),
\end{equation}
where $U$ obeys the uniform distribution in $[0,1]$.

\section{Derivation of the aging EAMSD (\ref{a})}\label{App0}

The integrand of (\ref{a}) can be rewritten as
\begin{equation}\label{A1}
	\langle (x(t+\Delta)-x(t))^2\rangle=\langle x^2(t+\Delta)\rangle+\langle x^2(t)\rangle
	-2\langle x(t+\Delta)x(t)\rangle,
\end{equation}
where the first two terms are known in (\ref{contion}), and the last term can be evaluated as
\begin{eqnarray}\label{A2}
		\langle x(t+\Delta)x(t)\rangle
		&=\langle x^2(t)\rangle+\int_{t}^{t+\Delta}\int_0^{t}
		\langle v(t_1)v(t_2)\rangle {\rm d}t_1{\rm d}t_2  \nonumber\\
		&=\langle x^2(t)\rangle+\nu\tau\int_{t}^{t+\Delta}\int_0^{t}
		t_1^{\beta-1}{\rm e}^{-\frac{t_2-t_1}{\tau}} {\rm d}t_1{\rm d}t_2.
\end{eqnarray}
We first calculate the integral with respect to $t_2$ on interval $(t,t+\Delta)$, and obtain
\begin{equation}\label{A3}
\langle x(t+\Delta)x(t)\rangle=\langle x^2(t)\rangle+\nu\tau\left(1-{\rm e}^{-\frac{\Delta}{\tau}}\right)\int_0^{t}
t_1^{\beta-1}{\rm e}^{-\frac{\Delta-t_1}{\tau}} {\rm d}t_1.
\end{equation}
It holds that
\begin{equation}\label{A4}
	\int_0^{t}
	t_1^{\beta-1}{\rm e}^{-\frac{\Delta-t_1}{\tau}} {\rm d}t_1
	\simeq\tau t^{\beta-1}
\end{equation}
for large $t$, since the Laplace transform $(t\rightarrow \lambda)$ of the left-hand side in (\ref{A4}) is $\frac{\Gamma(\beta)\lambda^{-\beta}}{\lambda+1/\tau}$, which asymptotically equals $\tau\Gamma(\beta)\lambda^{-\beta}$ for small $\lambda$. Combining (\ref{A1}), (\ref{A3}), (\ref{A4}), and (\ref{contion}), we can obtain
\begin{eqnarray}
	\langle (x(t+\Delta)-x(t))^2\rangle
	&\simeq \frac{2\nu\tau^2}{\beta}((t+\Delta)^\beta-t^\beta)
	  -2\nu\tau^3\left(1-{\rm e}^{-\frac{\Delta}{\tau}}\right) t^{\beta-1} \nonumber\\
	&\simeq 2\nu\tau^2t^{\beta-1}\left(\Delta-\tau\left(1-{\rm e}^{-\frac{\Delta}{\tau}}\right)\right)
\end{eqnarray}
for large $t$ and $\Delta\ll t$.
	
\section{Derivation of the velocity correlation function (\ref{ve})}\label{App1}

In the expression of the conditional correlation function $\langle v(t_1)v(t_2)|v_0,\tau\rangle$ (\ref{covar}), one finds that the terms containing parameter $\tau$ have the exponential form ${\rm e}^{-\frac{f}{\tau}}$.
By use of the notation $C=\frac{\alpha}{\Gamma(1/\alpha)\langle\tau\rangle}$
and the integral representation of the L\'{e}vy distribution
\begin{equation}
	L_\alpha(z)=\frac{1}{\alpha x}\frac{1}{2\pi {\rm i}}\int_{\gamma-{\rm i}\infty}^{\gamma+{\rm i}\infty}\frac{\Gamma(s/\alpha)}{\Gamma(s)}z^s {\rm d}s,~~~0<\alpha<1,
\end{equation}
the ensemble average on the exponential term ${\rm e}^{-\frac{f}{\tau}}$ over parameter $\tau$ is
\begin{eqnarray*}
    \int_0^\infty {\rm e}^{-\frac{f}{\tau}}g(\tau){\rm d}\tau
    &=\frac{\alpha}{\Gamma(1/\alpha)}\int_0^\infty {\rm e}^{-\frac{f}{\tau}}\frac{1}{\tau} L_\alpha(C\tau){\rm d}\tau \nonumber\\
    &=\frac{1}{\Gamma(1/\alpha)}\frac{1}{2\pi {\rm i}}\int_0^\infty {\rm e}^{-\frac{f}{\tau}}\frac{1}{\tau} \int_{\gamma-{\rm i}\infty}^{\gamma+{\rm i}\infty}\frac{\Gamma(s/\alpha)}{\Gamma(s)}(C\tau)^{s-1}{\rm d}s{\rm d}\tau\nonumber\\
    &=\frac{1}{\Gamma(1/\alpha)}\frac{1}{2\pi {\rm i}}\int_{\gamma-{\rm i}\infty}^{\gamma+{\rm i}\infty}\frac{\Gamma(s/\alpha)}{\Gamma(s)}
   (Cf)^{s-1} \int_0^\infty {\rm e}^{-\zeta}\zeta^{-s}{\rm d}\zeta {\rm d}s\nonumber\\
    &=\frac{1}{\Gamma(1/\alpha)}\frac{1}{2\pi {\rm i}}\int_{\gamma-{\rm i}\infty}^{\gamma+{\rm i}\infty}\frac{\Gamma(s/\alpha)\Gamma(1-s)}{\Gamma(s)}(Cf)^{s-1} {\rm d}s\nonumber\\
   &=\frac{\alpha}{\Gamma(1/\alpha)}\frac{1}{2\pi {\rm i}}\int_{\gamma-{\rm i}\infty}^{\gamma+{\rm i}\infty}\frac{\Gamma(s/\alpha+1)\Gamma(1-s)}{\Gamma(s+1)}(Cf)^{s-1} {\rm d}s.
\end{eqnarray*}
Denote the above integration as $R(f)$. It can be solved by using the residue theorem with poles $s/\alpha+1=-n$ or $1-s=-n$, where $n=0,1,2,...$.

In detail, using the poles $s=-(n+1)\alpha$ at the negative half plane, it holds that
\begin{eqnarray*}
 \fl
   R(f)
   =\frac{\alpha}{\Gamma(1/\alpha)}\sum_{n=0}^\infty \lim_{s\to -(n+1)\alpha}(s+(n+1)\alpha)\frac{\Gamma(s/\alpha+1)\Gamma(1-s)}{\Gamma(s+1)}
   (Cf)^{s-1}\\
  \fl =\frac{\alpha}{\Gamma(1/\alpha)}\sum_{n=0}^\infty \lim_{s\to -(n+1)\alpha}\alpha(s/\alpha+n+1)\frac{\Gamma(s/\alpha+2)\Gamma(1-s)}{(s/\alpha+1)\Gamma(s+1)}
   (Cf)^{s-1}\\
  \fl =\frac{\alpha}{\Gamma(1/\alpha)}\sum_{n=0}^\infty \lim_{s\to -(n+1)\alpha}\alpha(s/\alpha+n+1)\frac{\Gamma(s/\alpha+3)\Gamma(1-s)}{(s/\alpha+1)(s/\alpha+2)\Gamma(s+1)}
   (Cf)^{s-1}\\
  \fl =\frac{\alpha}{\Gamma(1/\alpha)}\sum_{n=0}^\infty \lim_{s\to -(n+1)\alpha}\alpha(s/\alpha+n+1)\frac{\Gamma(s/\alpha+n+2)\Gamma(1-s)(Cf)^{s-1}}{(s/\alpha+1)(s/\alpha+2)\cdot\cdot\cdot(s/\alpha+n+1)\Gamma(s+1)}
   \\
  \fl =\frac{\alpha}{\Gamma(1/\alpha)}\sum_{n=0}^\infty \lim_{s\to -(n+1)\alpha}\alpha\frac{\Gamma(s/\alpha+n+2)\Gamma(1-s)}{(s/\alpha+1)(s/\alpha+2)\cdot\cdot\cdot(s/\alpha+n)\Gamma(s+1)}
   (Cf)^{s-1}\\
  \fl =\frac{\alpha^2}{\Gamma(1/\alpha)}\sum_{n=0}^\infty \frac{(-1)^n}{n!}\frac{\Gamma(1+(n+1)\alpha)}{\Gamma(1-(n+1)\alpha)}(Cf)^{-(n+1)\alpha-1}.
\end{eqnarray*}
For large $f$, the asymptotic form of the above equation is
\begin{eqnarray}\label{large}
   R(f)
   \simeq\frac{\alpha^2}{\Gamma(1/\alpha)}\frac{\Gamma(1+\alpha)}{\Gamma(1-\alpha)}\left(\frac{f}{\tilde{\tau}}\right)^{-\alpha-1}.
\end{eqnarray}
Similarly, using the poles $s=n+1$ at the positive half plane, one can also obtain the expression of $R(f)$:
\begin{eqnarray}
   R(f)
   =\frac{1}{\Gamma(1/\alpha)}\sum_{n=0}^\infty \frac{(-1)^n}{n!}\frac{\Gamma((n+1)/\alpha)}{\Gamma(n+1)}\left(\frac{f}{\tilde{\tau}}\right)^{n}.
\end{eqnarray}
Substituting the asymptotic expression (\ref{large}) of $R(f)$ with $f=t_1+t_2$ and $f=t_1+t_2-2t'$  into the second line of (\ref{covar}), one can obtain the velocity correlation function averaging over parameter $\tau$ with large times $t_1$, $t_2$ and $t_1< t_2$
\begin{eqnarray}\label{vv}
 \fl
    \langle v(t_1)v(t_2)\rangle=\int_0^\infty\langle v(t_1)v(t_2)|v_0, \tau\rangle g(\tau){\rm d}\tau\nonumber\\
     \fl\simeq C_1
     \left(v_0^2 (t_1+t_2)^{-\alpha-1}
    +2\nu \int_0^{t_1}(t_1+t_2-2t')^{-\alpha-1}t'^{\beta-1} {\rm d}t'\right)\nonumber\\
     \fl=C_1
    \left(v_0^2 (t_1+t_2)^{-\alpha-1}
    + \frac{2\nu t_1^{\beta}(t_1+t_2)^{-\alpha-1}}{\beta} {_2F_1\left(\alpha+1,\beta,1+\beta,\frac{2t_1}{t_1+t_2}\right)}\right),
\end{eqnarray}
which is the velocity correlation function in (\ref{ve}) with $C_1=\frac{\Gamma(1+\alpha)}{\Gamma(1-\alpha)}\frac{\langle \tau\rangle^{1+\alpha}\Gamma(1/\alpha)^\alpha}{\alpha^{\alpha-1}}$.

\section{Derivation of the EAMSD (\ref{E})}\label{App2}

We only present the derivation of the second term of the EAMSD (\ref{E}), which is denoted as $\langle x^2(t)\rangle_2$ here. For long times, using the integral representation of the second term in velocity correlation function (\ref{vv}), i.e.,
\begin{eqnarray*}
 \langle v(t_1)v(t_2)\rangle_2=2\nu\frac{\Gamma(1+\alpha)}{\Gamma(1-\alpha)}\frac{\langle \tau\rangle^{1+\alpha}\Gamma(1/\alpha)^\alpha}{\alpha^{\alpha-1}}
     \int_0^{t_1}(t_1+t_2-2t')^{-\alpha-1}t'^{\beta-1} {\rm d}t',
\end{eqnarray*}
we have
\begin{eqnarray*}
 \fl
 \langle x^2(t)\rangle_2= 2\int_0^t\int_0^{t_2}\langle v(t_1)v(t_2)\rangle_2 {\rm d}t_1{\rm d}t_2\\
 \fl =4\nu\frac{\Gamma(1+\alpha)}{\Gamma(1-\alpha)}\frac{\langle \tau\rangle^{1+\alpha}\Gamma(1/\alpha)^\alpha}{\alpha^{\alpha-1}}
 \int_0^t\int_0^{t_2}\int_0^{t_1}(t_1+t_2-2t')^{-\alpha-1}t'^{\beta-1} {\rm d}t' {\rm d}t_1{\rm d}t_2\\
 \fl =2^{2-\beta}\nu\frac{\Gamma(1+\alpha)}{\Gamma(1-\alpha)}\frac{\langle \tau\rangle^{1+\alpha}\Gamma(1/\alpha)^\alpha}{\alpha^{\alpha-1}}
 \int_0^t\int_0^{t_2}(t_1+t_2)^{\beta-\alpha-1}\int_0^{\frac{2t_1}{t_1+t_2}} x^{\beta-1}(1-x)^{-\alpha-1}{\rm d}x{\rm d}t_1{\rm d}t_2\\
 \fl =2^{2-\beta}\nu\frac{\Gamma(1+\alpha)}{\Gamma(1-\alpha)}\frac{\langle \tau\rangle^{1+\alpha}\Gamma(1/\alpha)^\alpha}{\alpha^{\alpha-1}}
 \int_0^t t_2^{\beta-\alpha}\int_0^1 (\eta+1)^{\beta-\alpha-1}\int_0^{\frac{2\eta}{\eta+1}} x^{\beta-1}(1-x)^{-\alpha-1}{\rm d}x{\rm d}\eta {\rm d}t_2\\
 \fl =\frac{\Gamma(1+\alpha)}{\Gamma(1-\alpha)}\frac{\langle \tau\rangle^{1+\alpha}\Gamma(1/\alpha)^\alpha}{\alpha^{\alpha-1}} \frac{2^{2-\beta}\nu F}{1-\alpha+\beta}t^{1-\alpha+\beta},
\end{eqnarray*}
where we have used the variable substitution $\frac{2t'}{t_1+t_2}=x$ in the third line and $\frac{t_1}{t_2}=\eta$ in the fourth line. Besides,
$F=\int_0^1 (\eta+1)^{\beta-\alpha-1}\int_0^{\frac{2\eta}{\eta+1}} x^{\beta-1}(1-x)^{-\alpha-1}{\rm d}x{\rm d}\eta$ is a constant.

\section{The evaluations of two terms in (\ref{TAMSDDivide})}\label{App3}
Denote the two terms in (\ref{TAMSDDivide}) as $I_1$ and $I_2$, respectively. For $I_1$, we have
\begin{equation}\label{S3-1}
  I_1=2\nu\Delta\int_0^\Delta \tau^2 g(\tau){\rm d}\tau.
\end{equation}
Substituting the expression (\ref{PDF}) of $g(\tau)$  into (\ref{S3-1}), and performing the variable substitution yield
\begin{equation*}
I_1=\frac{2\nu\Delta \langle\tau\rangle}{C}\int_0^{C\Delta}
\tau L_\alpha(\tau){\rm d}\tau,
\end{equation*}
the large $\Delta$ asymptotics of which can be evaluated by
using Laplace transform. The Laplace transform ($C\Delta\rightarrow \lambda$) of the integral is
\begin{equation*}
  -\frac{1}{\lambda}\frac{{\rm d}}{{\rm d}\lambda}{\rm e}^{-\lambda^\alpha}\simeq \alpha\lambda^{\alpha-2}
\end{equation*}
for small $\lambda$, and its inverse Laplace transform is $\frac{\alpha (C\Delta)^{1-\alpha}}{\Gamma(2-\alpha)}$. Therefore,
\begin{equation*}
I_1\simeq \frac{2\nu\alpha\langle\tau\rangle}{C^\alpha\Gamma(2-\alpha)}\Delta^{2-\alpha}.
\end{equation*}

Similarly, $I_2$ can be calculated as
\begin{eqnarray*}
	I_2=\nu\Delta^2\int_\Delta^\infty\tau g(\tau){\rm d}\tau
	=\nu\Delta^2\langle\tau\rangle\int_{C\Delta}^\infty L_\alpha(\tau){\rm d}\tau \simeq\frac{\nu\langle\tau\rangle}{\Gamma(1-\alpha)C^{\alpha}}\Delta^{2-\alpha},
\end{eqnarray*}
where we evaluate the integral in second equality by using Laplace transform ($C\Delta\rightarrow \lambda$), i.e.,
\begin{equation*}
\frac{1}{\lambda}-\frac{1}{\lambda}{\rm e}^{-\lambda^\alpha}\simeq\lambda^{\alpha-1},
\end{equation*}
the inverse Laplace transform of which is $\frac{(C\Delta)^{-\alpha}}{\Gamma(1-\alpha)}$.
Therefore, based on the approximation in (\ref{TAMSDDivide}), the ensemble-averaged TAMSD can be obtained as
\begin{eqnarray*}
	\langle \overline{{\delta^2(\Delta)}}\rangle
	\simeq I_1+I_2
	\simeq \frac{\nu(1+\alpha)\langle\tau\rangle}{(C)^\alpha\Gamma(2-\alpha)}\Delta^{2-\alpha},
\end{eqnarray*}
the coefficient of which is approximately equal to the one of (\ref{EA-TA-beta1}).

\section*{References}
\bibliographystyle{iopart-num}
\bibliography{ReferenceCW}

\end{document}